\documentclass[a4paper,fleqn,usenatbib]{mnras}

\usepackage{caption}
\usepackage[final]{graphicx}	
\usepackage{amsmath}	
\usepackage{amssymb}	
\usepackage{multicol}        
\usepackage{bm}		
\usepackage{pdflscape}	

\usepackage[T1]{fontenc}
\usepackage{ae,aecompl}

\usepackage{graphicx}	
\usepackage{amsmath}	
\usepackage{amssymb}	

\title[Discarding orbital decay in WASP-19b]{Discarding orbital decay in WASP-19b after one decade of transit observations\thanks{This work is partially based on observations obtained with the 1.54-m telescope at Estaci\'on Astrof\'isica de Bosque Alegre dependent on the National University of C\'ordoba, Argentina.}\thanks{Based on data acquired at Complejo Astron\'omico El Leoncito, operated under agreement between the Consejo Nacional de Investigaciones Cient\'ificas y T\'ecnicas de la Rep\'ublica Argentina and the National Universities of La Plata, C\'ordoba and San Juan (Programme ID: JS-2019A-06, PI: E. Jofr\'e).}}

\author[R. Petrucci et al.]{R. Petrucci$^{1,2,6}$\thanks{E-mail: romina@astro.unam.mx}, E. Jofr\'e$^{1,2,6}$, Y. G\'omez Maqueo Chew$^{1}$, T. C. Hinse$^{3}$, M. Ma\v{s}ek$^{4}$,
\newauthor{T.-G. Tan$^{5}$, and M. G\'omez$^{2,6}$}
\\
$^{1}$Instituto de Astronom\'{i}a, Universidad Nacional Aut\'{o}noma de M\'{e}xico, Ciudad Universitaria, 04510 Ciudad de M\'{e}xico, M\'{e}xico\\
$^{2}$Universidad Nacional de C\'{o}rdoba, Observatorio Astron\'{o}mico, Laprida 854, 5000 C\'{o}rdoba, Argentina \\
$^{3}$Chungnam National University, Department of Astronomy and Space Science, 34134 Daejeon, Republic of Korea\\
$^{4}$Institute of Physics Czech Academy of Sciences, Na Slovance 1999/2, CZ-182 21 Praha, Czech Republic\\
$^{5}$Perth Exoplanet Survey Telescope, Perth, Australia\\
$^{6}$Consejo Nacional de Investigaciones Cient\'ificas y T\'ecnicas (CONICET), Argentina
}

\date{Accepted XXX. Received YYY; in original form ZZZ}

\pubyear{2015}

\begin{document}
\label{firstpage}
\pagerange{\pageref{firstpage}--\pageref{lastpage}}
\maketitle

\begin{abstract}
We present a empirical study of orbital decay for the exoplanet WASP-19b, based on mid-time measurements of 74 complete transits (12 newly obtained by our team and 62 from the literature), covering a 10-year baseline. A linear ephemeris best represents the mid-transit times as a function of epoch. Thus, we detect no evidence of the shortening of WASP-19b's orbital period and establish an upper limit of its steady changing rate, $\dot{P}=-2.294$ ms $yr^{-1}$, and a lower limit for the modified tidal quality factor $Q'_{\star} = (1.23 \pm 0.231) \times 10^{6}$. Both are in agreement with previous works. This is the first estimation of $Q'_{\star}$ directly derived from the mid-times of WASP-19b obtained through homogeneously analyzed transit measurements.
Additionally, we do not detect periodic variations in the transit timings within the measured uncertainties in the mid-times of transit. We are therefore able to discard the existence of planetary companions in the system down to a few $M_\mathrm{\earth}$ in the first order mean-motion resonances 1:2 and 2:1 with WASP-19b, in the most conservative case of circular orbits.
Finally, we measure the empirical $Q'_{\star}$ values of 15 exoplanet host stars which suggest that stars with $T_\mathrm{eff}$ $\lesssim$ 5600K dissipate tidal energy more efficiently than hotter stars. This tentative trend needs to be confirmed with a larger sample of empirically measured $Q'_{\star}$.\end{abstract}




\section{Introduction}

WASP-19b is a short-period gas giant planet and one of the most well characterized so far. Since its discovery \citep{hebb}, the stellar and planetary properties of the system have been re-determined by several authors \citep{hellier, dragomir, bean, tregloan, lendl, mancini, seda, espinoza}. The results of these studies confirm that WASP-19b is a Jupiter-like planet ($M_\mathrm{P}=$ 1.11 $M_\mathrm{J}$, $R_\mathrm{P}=$ 1.39 $R_\mathrm{J}$) in a very short orbit (P $\sim$ 19 h) hosted by an active G8V solar-type star ($M_\mathrm{\star}=$ 0.9 $M_\mathrm{\sun}$, $R_\mathrm{\star}=$ 1.0 $R_\mathrm{\sun}$, Age $\sim$ 11 Gyr; \citealt{hebb, knutson, anderson, huitson, tregloan}). Given the brief time it takes to orbit once around its host, WASP-19b was one of the first ultra-short period planets discovered.  

Planets so close to their host stars are subject to strong tidal interactions that may entirely dominate the planetary orbital evolution \citep{levrard, matsumura}. Particularly, in systems such as WASP-19, where the stellar rotation period is larger than the orbital period, the tidal dissipation in the star might transfer angular momentum from the orbit to the stellar spin, leading to orbital decay. According to the calculations performed by \citet{matsumura}, because the total angular momentum of the system (considering conservatively that it is conserved) is smaller than the critical value (which depends on the masses and moments of inertia of both star and planet), the planetary orbit is unstable and the planet is expected to cross its Roche limit and to be tidally disrupted. By considering convective damping equilibrium tides as prescribed by \citet{zahn77}, \citet{valsechi} predicted that as a result of orbital decay after 10 years of observations the transit arrival time of WASP-19b would be shifted by a minimum of 34 seconds. \citet{essick} studied the orbital evolution of several hot Jupiters due to the excitation of g-modes in solar-type stars as a consequence of the tides raised by the planets. For the WASP-19 system, they found a decay time of 9.2 Myr and a mid-transit time shift of 191 seconds after 10 years. Moreover, recently, the results of some empirical studies \citep{mancini, espinoza}, based on transit observations, have provided hints of a possible variation in the mid-transit times of WASP-19b. In this sense, \citet{mancini} found that a linear ephemeris is not a good fit ($\chi^{2}_\mathrm{r}=$ 1.98) for the 54 transit times that they analyzed as a function of the epoch, although these authors attribute it to a subestimation of the transit time errors. On the other hand, \citet{espinoza} computed new ephemeris for the system by considering the 54 datasets of \citet{mancini} and their 6 new transit light curves. They detected a systematic difference of $\sim$ 40 seconds between the 2014-2015 data and those of 2017. Given that the errors in these transit timings are 10 seconds at most, the authors interpret this difference as significant.     
In order to elucidate if the orbit of WASP-19b is suffering measurable orbital decay, in 2016 we started a photometric monitoring of the primary transits of this system, that produced a total of 12 complete transit light curves. In this work, we present the results of the homogeneous analysis of these newly acquired transits and 62 transit light curves from the literature, that allowed us to study the long-term evolution of the system over one decade and test the predictions of its expected orbital decay.

This paper is organized as follows: in Section \ref{observations}, we present our observations and data reduction. In Section \ref{section:3}, we describe the data modeling and the determination of the photometric parameters. The analysis of how the mid-transit times are affected by the selection of the initial parameters is presented in Section \ref{section:4}, and the study of orbital decay and periodic transit timing variations, including the search for companions in the system and a dynamical analysis, in Section \ref{section:5}. Finally, the summary and discussion are in Section \ref{section:6}. 

\section{OBSERVATIONS AND DATA REDUCTION}\label{observations}

We observed 12 new full transits of WASP-19b between May 2016 and April 2019 with the four different facilities described below.

\subsection{CASLEO Observatory}

Three complete transit light curves were obtained with the 2.15-m \textit{Jorge Sahade} telescope at CASLEO (Complejo Astron\'omico El Leoncito) in Argentina. All the observations were performed with the 2048$\times$2048 13.5 $\mu$m-size pixel \textit{Roper Scientific} camera, with a circular FoV of 9' diameter provided by a focal reducer at a plate scale of 0.45 arcsec per pixel. Two transits were observed with a Johnson R filter and the other one with a Johnson V filter. As calibration images, we took 10 bias, 10 sky- and 10 dome-flat frames per night. Dark frames were not acquired due to a low intensity dark current level. In all the three cases, an averaged bias was subtracted from the science images acquired during the night and then, they were divided by a master flat obtained as the median combined bias-corrected individual sky-flats using standard IRAF\footnote{IRAF is distributed by the National Optical Astronomy Observatories, which are operated by the Association of Universities for Research in Astronomy, Inc., under cooperative agreement with the National Science Foundation.} routines. The lowest dispersions in relative flux were computed for the transit light curves obtained from images corrected by sky-flat field frames, so we used them (and not the ones coming from images corrected by dome-flat fields) for the analysis presented in the next sections. 

\subsection{EABA Observatory}

Six complete transit light curves were acquired with the 1.54-m telescope, operated with the Newtonian focus, at the Estaci\'on Astrof\'isica de Bosque Alegre (EABA) in Argentina. The observations of the first five transits were carried out with an Apogee Alta F16M camera of 4096$\times$4096 9 $\mu$m-size pixel with a FoV=16.8'$\times$16.8' and a scale of 0.25 arcsec per pixel. While the observations of the most recent transit were performed with a 3070$\times$2048, 9 $\mu$m-size pixel Apogee Alta U9 camera, which also provides a scale of 0.25 arcsec per pixel and a 8'$\times$12' field of view. We used a Johnson R filter for the observation of five transits and a Johnson I filter for the other one. As calibration images, each night we took 10 bias, 8 dark frames and 15 dome flat-fields in the corresponding band. By using standard IRAF routines, we subtracted from science images an averaged bias and a median-combined bias-corrected dark frame and finally divided them by a master flat generated as the median combined bias- and dark-corrected flat-field frames.

\subsection{PEST Observatory}

Two transits (one complete and one partial) were observed in the R$_\mathrm{C}$ band at the Perth Exoplanet Survey Telescope (PEST) observatory located in Western Australia. PEST is a home observatory with a 12-inch Meade LX200 SCT f/10 telescope with an SBIG ST-8XME CCD camera, and is equipped with a BVR$_\mathrm{C}$I$_\mathrm{C}$ filter wheel, a focal reducer yielding f/5, and an Optec TCF-Si focuser. PEST has a 31'$\times$21' field of view and a 1.2 arcsec per pixel scale. Image calibration was done with master darks and twilight flats, consisting of 40 and $\sim$ 100 individual frames respectively. The image reduction was done using C-Munipack\footnote{http://c-munipack.sourceforge.net/}.

\subsection{FRAM Telescope}

Two complete transit light curves were observed with the FRAM (F/(Ph)otometric Robotic Atmospheric Monitor) telescope, which is a part of the Pierre Auger Observatory located near the town of Malarg\"ue in the province of Mendoza, Argentina. The main task of the FRAM telescope is the continuous night-time monitoring of the atmospheric extinction and its wavelength dependence for the Pierre Auger Observatory. The additional activities of the FRAM telescope include photometry of selected variable stars and exoplanets, astrometry measurement and photometry of comets and asteroids, and observations of optical counterparts of gamma ray bursts. FRAM is a 12-inch, f/10 Meade Schmidt-Cassegrain equipped with a Optec 0.66$\times$ focal reducer and a micro-focuser. The light from the telescope is collected by a Moravian Instruments CCD camera G2-1600 with the KAF-1603ME sensor that has 1536$\times$1024 pixels and a field of view of 23'$\times$15'. The camera has a maximum quantum efficiency of more than 80$\%$ and an integrated filter wheel occupied by a set of photometric BVRI filters. This setup reaches 16-17 mag for a 60 second unfiltered exposure.\\

Middle times of CASLEO, EABA, and PEST images were recorded in Heliocentric Julian Date based on Coordinated Universal Time ($\mathrm{HJD_\mathrm{UTC}}$), while those of the FRAM images are in units of Julian Date based on Coordinated Universal Time ($\mathrm{JD}_\mathrm{UTC}$).

On the reduced images acquired at the CASLEO, EABA, and FRAM facilities, we measured precise instrumental magnitudes with the FOTOMCAp code \citep{petrucci16}. This code applies the method of aperture correction \citep{howell, stetson}, which combines the magnitude measured with the growth-curve technique and the one determined at the aperture that provides the highest signal-to-noise ratio for every star at each individual image. The photometric error adopted for each instrumental magnitude is that computed by the IRAF \textit{phot} task. 
After transforming the instrumental magnitudes of all the stars found by the IRAF \textit{daofind} task in fluxes, we obtained the transit light curve of each night by dividing the flux of WASP-19, measured per image, by the summed fluxes of an ensemble of non-variable stars with the same brightness as the science target, when possible. For the reduced images taken at PEST Observatory, the photometry was done by using C-Munipack adopting the same criteria explained before to select the comparison stars. Specific details of the observations of our complete transits are presented in Table \ref{table:1}.  

\begin{table*}
\caption{Log of our observations}             
\label{table:1}      
\centering          
\begin{tabular}{lcccccccc}     
\hline\hline       
Date & Telescope & Camera & Filter & Bin-size & X & Exposure Time (s) & N$_{\mathrm{obs}}$ & $\sigma$(mag)\\
\hline                    
2016 May 09 & 0.30-m PEST & SBIG & $R_\mathrm{C}$ & 1x1 & 1.0 $\rightarrow$ 2.4  & 120 & 117 & 0.0022\\
2016 Jun 06 & 1.54-m EABA & F16M & $R$ & 2x2 & 1.23 $\rightarrow$ 2.36 & 40 & 310 & 0.0025\\
2016 Jun 10 & 1.54-m EABA & F16M & $R$ & 2x2 & 1.09 $\rightarrow$ 2.0 & 40 & 302 & 0.0033\\
2016 Jul 06 & 0.30-m FRAM & G2-1600  & $R$ & 1x1 & 1.28 $\rightarrow$ 2.79 & 120 & 91 & 0.0034\\
2016 Dec 16 & 1.54-m EABA & F16M & $R$ & 2x2 & 1.17 $\rightarrow$ 1.03 & 25, 35 & 432 & 0.0026\\
2017 Jan 23 & 2.15-m CASLEO & Roper & $R$ & 2x2 & 1.36 $\rightarrow$ 1.04 & 12, 15, 17, 30, 40 & 461 & 0.0026\\
2017 Jan 30 & 1.54-m EABA & F16M & $I$ & 2x2 & 1.06 $\rightarrow$ 1.25 & 50 & 276 & 0.0018\\
2017 Feb 07 & 0.30-m FRAM & G2-1600 & $R$ & 1x1 & 1.16 $\rightarrow$ 1.01 & 120 & 83 & 0.0036\\
2017 Nov 21 & 1.54-m EABA & F16M & $R$ & 4x4 & 1.65 $\rightarrow$ 1.08 & 20, 30 & 376 & 0.0033\\
2017 Dec 21 & 2.15-m CASLEO & Roper & $V$ & 2x2 & 1.38 $\rightarrow$ 1.04 & 30 & 281 & 0.0029\\
2019 Feb 12 & 2.15-m CASLEO & Roper & $R$ & 2x2 & 1.16 $\rightarrow$ 1.73 & 27 & 196 & 0.0035\\
2019 Apr 10 & 1.54-m EABA & U9 & $R$ & 2x2 &  1.05 $\rightarrow$ 1.67 & 40 & 449 & 0.0017\\ 
\hline                  
\end{tabular}

Note: Date is given for the beginning of the transit, X is the
airmass change during the observation, N$_{\mathrm{obs}}$ is the number of useful exposures, and $\sigma$ is the standard deviation of the out-of-transit data-points.
\end{table*}

\subsection{Literature and public data}

Given that the main purpose of this study is to analyze the mid-transit times evolution of WASP-19b over the longest possible time-coverage, we collected other 62 complete phase-coverage transit light curves from the previous works and public databases listed below:\\

\begin{itemize}

\item 
\cite{hebb}: One transit observed in the z-band with the 2-m Faulkes Telescope South (FTS) in Australia on 2008 December 17. The images central times were recorded in $\mathrm{HJD_\mathrm{UTC}}$.\\

\item
\cite{hellier}: One transit observed in a Gunn-r filter with the 3.58-m New Technology Telescope (NTT), operated at European Southern Observatory (ESO) La Silla in Chile on the night of 2010 February 28. The images central times were recorded in $\mathrm{HJD_\mathrm{UTC}}$.\\

\item
\cite{dragomir}: One transit observed in a Cousins R-band filter with the 1-m telescope at Cerro Tololo Inter-American Observatory (CTIO) in Chile on the night of 2011 January 18. The central times of the images are in units of $\mathrm{JD_\mathrm{UTC}}$.\\

\item
\cite{lendl}: A total of 11 transits observed between May 2010 and May 2012. Three of them obtained in the IC-, Gunn-z'-, and Gunn-r'-bands with the EulerCam at the 1.2-m Euler-Swiss Telescope at ESO La Silla Observatory (Chile), and the eight remaining transits observed in the I+z' filter with the 0.6-m TRAPPIST telescope, also at ESO La Silla Observatory in Chile. All the middle times in the images are in $\mathrm{HJD_\mathrm{UTC}}$.\\

\item
\cite{mancini}: A total of 10 transits observed between May 2010 and April 2012. Seven of them were obtained in a Gunn-i filter using the DFOSC imager mounted on the 1.54-m Danish Telescope at La Silla in Chile and the other three, observed simultaneously on the night of 2012 April 15 in the J-, K-, and H-band with the Gamma Ray Burst Optical and Near-Infrared Detector (GROND) instrument mounted on the MPG/ESO 2.2-m telescope also at ESO La Silla. The images central times were recorded in $\mathrm{HJD_\mathrm{UTC}}$.\\

\item
\cite{bean}: 18 light curves corresponding to two transits obtained in white light through multi-object spectroscopy the nights of 2012 March 13 and 2012 April 4 with the MMIRS instrument on the Magellan II (Clay) telescope at Las Campanas Observatory in Chile. Each transit observed in white light was split in nine light curves obtained at different wavelength bins, ranging from 1.25 to 2.35 $\mu$m. In this case, central times were recorded in Heliocentric Julian Date in Barycentric Dynamical Time ($\mathrm{HJD_\mathrm{TDB}}$).\\

\item
\cite{seda}: three transits observed between November 2014 and February 2016 through multi-object spectroscopy with the 8.2-m Unit Telescope 1 (UT1) of the ESO's Very Large Telescope (VLT) in Chile. For each observation, they used the grisms: 600B (blue), 600RI (green) and 600z (red) of the FORS2 spectrograph, which cover the wavelength domain between 0.43 and 1.04 $\mu$m. In this case, the available data consist of three tables with times and normalized fluxes measured on a series of wavelength bins spanning the full wavelength range in which each transit was observed. We recovered the three light curves i.e., estimated the flux and error for each time of every night, by performing the average and standard deviation of the normalized fluxes estimated in each bin. The central times of the images are in units of $\mathrm{JD_\mathrm{UTC}}$.\\

\item
\cite{espinoza}: three transits observed in white light through multi-object spectroscopy between March 2014 and April 2017 with the Inamori-Magellan Areal Camera and Spectrograph (IMACS), mounted at the Magellan Baade 6.5-m Telescope in Las Campanas Observatory (LCO) in Chile. Central times of the images are in units of Barycentric Julian Dates in Barycentric Dynamical Time ($\mathrm{BJD}_\mathrm{TDB}$). Given that the photometric errors were not provided, we adopted the standard deviation of the out-of-transit data points as the uncertainty of each measured magnitude in the light curve.\\ 

\item
ETD (Exoplanet Transit Database\footnote{The Exoplanet Transit Database (ETD) can be found at:http://var2.astro.cz/ETD/credit.php; see also TRESCA at:http://var2.astro.cz/EN/tresca/index.php}; \citealt{poddany}): A total of 14 transits observed between March 2010 and March 2019, one obtained in a I filter, one in a V filter, one in an R filter, three in an R$_\mathrm{C}$ filter and the remaining ones without using filter. These transits were observed with telescopes of different sizes ranging from 0.23- to 2.15-m. For those light curves with no photometric errors, we adopted the standard deviation of the out-of-transit data points as uncertainty. The middle times were recorded in $\mathrm{HJD_\mathrm{TDB}}$ and in Geocentric Julian Date based on Coordinated Universal Time ($\mathrm{GJD_\mathrm{UTC}}$).\\ 

\end{itemize}

As further explained in Section \ref{section:5}, the mid-times of the transits collected from the literature, including all from the ETD, as well as those observed by our team were converted to the Barycentric Julian Date system \citep{eastman}.

\section{DETERMINATION OF PHOTOMETRIC PARAMETERS}\label{section:3}

We used version 34 of the JKTEBOP\footnote{http://www.astro.keele.ac.uk/jkt/codes/jktebop.html} code \citep{sou04} to fit the 74 full-transit light curves. This code assumes both, star and planet, are represented as biaxial spheroids (adopting the spherical approximation for the calculation of light lost during transit) and uses numerical integration of concentric circles over each component to calculate the resulting light curve. In order to determine the photometric parameters of the system, the fitted quantities were the inclination ($i$), the sum of the fractional radii ($\Sigma=r_{\star}+r_\mathrm{P}$)\footnote{$r_{\star} =\frac{R_{\star}}{a}$ and $r_\mathrm{P} =\frac{R_{P}}{a}$ are the ratios of the absolute radii of the star and the exoplanet, respectively, to the semimajor axis ($a$).}, the ratio of the fractional radii ($k = r_{P}/r_{\star}$), the flux level of the out-of-transit data ($l_{0}$), the linear and non-linear quadratic limb-darkening coefficients ($q_{1}$ and $q_{2}$), the mid-transit time ($T_\mathrm{0}$), and the coefficients of a second order polynomial to normalize the light curves\footnote{The simultaneous fitting of this second order polynomial enables to normalize the light curve by removing any parabolic trend produced by differential extinction, stellar activity, and/or differences between the spectral types of the stars used as comparisons and WASP-19. However, this approach is not good to eliminate red noise components associated with shifts in the position of the star on the CCD, seeing fluctuations, background variations, among others, that affect the measured stellar flux. In order to account for and properly remove this systematics, it would be required, for example, to multiply the transit model by a polynomial of any combination of these parameters with an order higher than two.}. The values of the orbital period ($P$), eccentricity ($e$), and mass ratio were kept as fixed quantities for our entire analysis.

As initial values for the parameters modeled by JKTEBOP, we considered the values of $i$, $\Sigma$, $k$, and $P$ estimated by \citet{mancini}, $e$ equal to zero, and mass ratio $= 0.00114$ calculated as the ratio of the planet and stellar masses computed by \citet{hellier}. Also, we assumed a quadratic limb darkening (LD) law, except for the transit light curves of \citet{bean} for which we adopted linear LD coefficients, indicated by the authors as the LD law that allows the best-fitting to the observations. 
For all the transits, the values of the linear and non-linear coefficients for WASP-19 were calculated with the JKTLD\footnote{http://www.astro.keele.ac.uk/jkt/codes/jktld.html} code by bilinearly interpolating $T_\mathrm{eff}$ and $\log g$ of the host star in published tables of coefficients calculated from stellar model atmospheres. In this case, we used the tabulations of \citet{claret00, claret04} computed from ATLAS atmospheric models \citep{kurucz}. As input for JKTLD, we adopted for WASP-19, $T_\mathrm{eff}=5591 \pm 62$ K and $\log g=4.46 \pm 0.09$ cgs as measured by \citet{mortier}, [Fe/H]$=0.3$ dex and v$_\mathrm{turb}=2$ $\mathrm{kms^{-1}}$.  

Given that many of the filters used to carry out the observations do not have theoretical LD coefficients computed in \citet{claret00, claret04}, we considered as initial LD values those corresponding to the filters with effective wavelengths closest to the real ones. Then, for the observations performed in the Cousins $I+$Sloan $z'$ and $z'$ bands, we adopted the values of the Sloan $z'$ filter. For transits acquired in the Johnson $R$ and $I$ bands, the values tabulated for the Cousins $R$ and $I$ filters were used. For the Gunn $r'$ band, the Sloan $r'$ filter was adopted. In the case of transits observed in the Gunn $i'$, the Sloan $i'$ filter was considered.  For those light curves observed with no filter, we used the average of the values in the Johnson $V$ and the Cousins $R$ bands as in \citet{nascimbeni}. Finally, the initial LD coefficients assumed for the observations in the J-, H-, and K- bands of GROND and those obtained by \citet{bean} in the same wavelength range, were those of Johnson J, H, and K. Here, it is important to notice that, as previously demonstrated in \citet{petrucci18}, the initial values adopted for the LD coefficients, even if they correspond to a filter with a different effective wavelength than that of the band used to perform the observations, cause a negligible effect on the measured mid-transit times with changes in the ephemeris within its 1$\sigma$ error.

As in previous works, once the initial values of the parameters were set, we performed a two steps procedure to find the best-fitting model for each transit:

\begin{itemize}

\item
First, we used a Levenberg-Marquardt optimization algorithm provided by JKTEBOP to carry out three different fits. Each fit regards as free parameters $i$, $\Sigma$, $k$, $l_{0}$, $T_\mathrm{0}$, the coefficients of a second order polynomial and also considers: a) the linear and non-linear LD coefficients as free quantities, or b) the linear coefficient slightly perturbed and the non-linear freely varying, or c) both LD coefficients fixed. After comparing the results given by these three possibilities, we adopted as the best model the one with the smallest value of $\chi^{2}_{r}$. The best model found indicates how the LD coefficients of each particular transit have to be treated (options a, b, or c) all along this section and also in Sections \ref{section:4} and \ref{section:5}, which include the influence of systematics on the measurement of mid-times and the determination of their values and errors. Before continuing with the next step, we multiplied the photometric uncertainties by the square root of the reduced chi-squared of the fit to get $\chi^{2}_{r}=1$.\\

\item
Second, we estimated realistic errors and mean values for the fitted parameters with two tasks provided by JKTEBOP: a residual permutation (RP) algorithm which accounts for the red noise in the photometric data and Monte Carlo simulations (10\,000 iterations). To be conservative, we kept the results obtained with the task that provided the largest error. Then, the median and the asymmetric uncertainties, $\sigma_{+}$ and $\sigma_{-}$, defined by the 15.87th and 84.13th percentiles values of the selected distribution (i.e., $-1 \sigma$ and $+1 \sigma$ respectively) were adopted as the best-fitting values and errors for the fitted parameters. 

\end{itemize}

In this study, we decided to analyze all the transits individually, instead of performing a joint analysis, to obtain the best set of photometric parameters of each transit independently fitted. Given that all the transits were homogeneously adjusted through the same fitting procedure, we estimated a new set of reliable values for the photometric parameters of the system from the best/high quality transits in our sample. To assess the quality of each transit light curve, we used two metrics: the photometric noise rate (PNR) and the $\beta$ factor.

The first one is defined by \citet{fulton} as,

\begin{equation}
      PNR = \frac{rms}{\sqrt{ \Gamma}},
      \label{eq:pnr}
\end{equation}

\noindent where $rms$ is the standard deviation of the transit residuals and $\Gamma$ represents the median number of exposures per minute. 
On the other hand, the $\beta$ factor or the red noise level is defined by \citet{winn08} as, 

 \begin{equation}
      \beta=\frac{\sigma_{\mathrm{r}}}{\sigma_{\mathrm{N}}},
      \label{eq:beta}
\end{equation}

\noindent where $\sigma_{\mathrm{N}}$ represents the expected standard deviation in the residuals, without binning, and $\sigma_{\mathrm{r}}$ is the standard deviation of the residual average values computed into M bins of N points each. In this work, the residuals were averaged in bins from 10 to 30 minutes and, through Eq. \ref{eq:beta}, a $\beta$ value was measured for each of them. The median of these individual measurements was considered as the red noise level for the light curve. In summary, PNR characterizes the scatter of the light curve (white noise) considering a specific time interval, while $\beta$ describes the degree of correlation among the data points (red noise).\\ 

After carefully inspecting the transit observations and their best fits, we considered the 40 light curves with PNR $\le$ 3 mmag and $\beta$ $\le$ 1.1 as the highest quality transits of our sample. For these selected light curves, we kept the median and asymmetric errors of the photometric parameters given by the algorithm (Monte Carlo or Residual Permutation) which gave the largest error. Then, following the standard procedure, the system's final values of $i$, $\Sigma$, and $k$ were calculated as the weighted average and the standard deviation of the measurements of each of the 40 chosen transits\footnote{This methodology assures that the final estimations of $i$, $\Sigma$, and $k$ are not biased by the values derived from low significance data which present the largest errors. This is owing to each of these final quantities is computed as the weighted average of the values measured only from the 40 high-quality transits of our sample, where the weights are calculated as the inverse of the quadratic error in the parameter.}. These values are in agreement, within errors, with the estimations of previous works \citep{hebb, lendl, tregloan, bean, mancini, espinoza} as can be seen in Table \ref{table:2}, where they are compared with the results of some of these past studies. Although our measurements are not as precise as the values reported in the literature, it is worthwhile to notice that they were computed from a large sample of transits all homogeneously analyzed, and we have conservatively estimated our uncertainties to avoid skewing our timing results.
 
\begin{table*}
\caption{Photometric parameters derived in this work along with the values previously determined by \citet{tregloan}, \citet{mancini}, and \citet{espinoza}}             
\label{table:2}      
\centering          
\begin{tabular}{lcccc}     
\hline\hline       
Parameter & This work	&	\citet{tregloan} & \citet{mancini}	& \citet{espinoza}\\
\hline                    
$i$	&	79.3 $\pm$ 1.3 &	78.94 $\pm$ 0.23	&   78.76 $\pm$ 0.13 & 79.29 $\pm$ 0.10 \\
$k$	&	0.145	$\pm$ 0.006	&	0.1428  $\pm$ 0.0006 & 0.14259 $\pm$ 0.00023 &	0.14233 $\pm$ 0.00050\\
$r_{\star}$+$r_\mathrm{P}$	& 0.3245 $\pm$ 0.0147	&	0.3301 $\pm$ 0.0019 & 0.33091 $\pm$ 0.00074 & --  \\
$r_{\star}$	& 0.2838 $\pm$ 0.0120 &	 --  &  0.28968 $\pm$ 0.00065  & 0.28169$^{a}$ $\pm$ 0.00111$^{a}$  \\
$r_\mathrm{P}$	& 0.039 $\pm$ 0.004	& -- & 0.04124 $\pm$ 0.00012 & --  \\
\hline                  

\end{tabular}

$^{a}$: These numbers were calculated from the value of $a/R_{\star}$ and its error
published in \citet{espinoza}.

\end{table*}

\begin{figure*}
  \centering
 \includegraphics[width=1.0\textwidth]{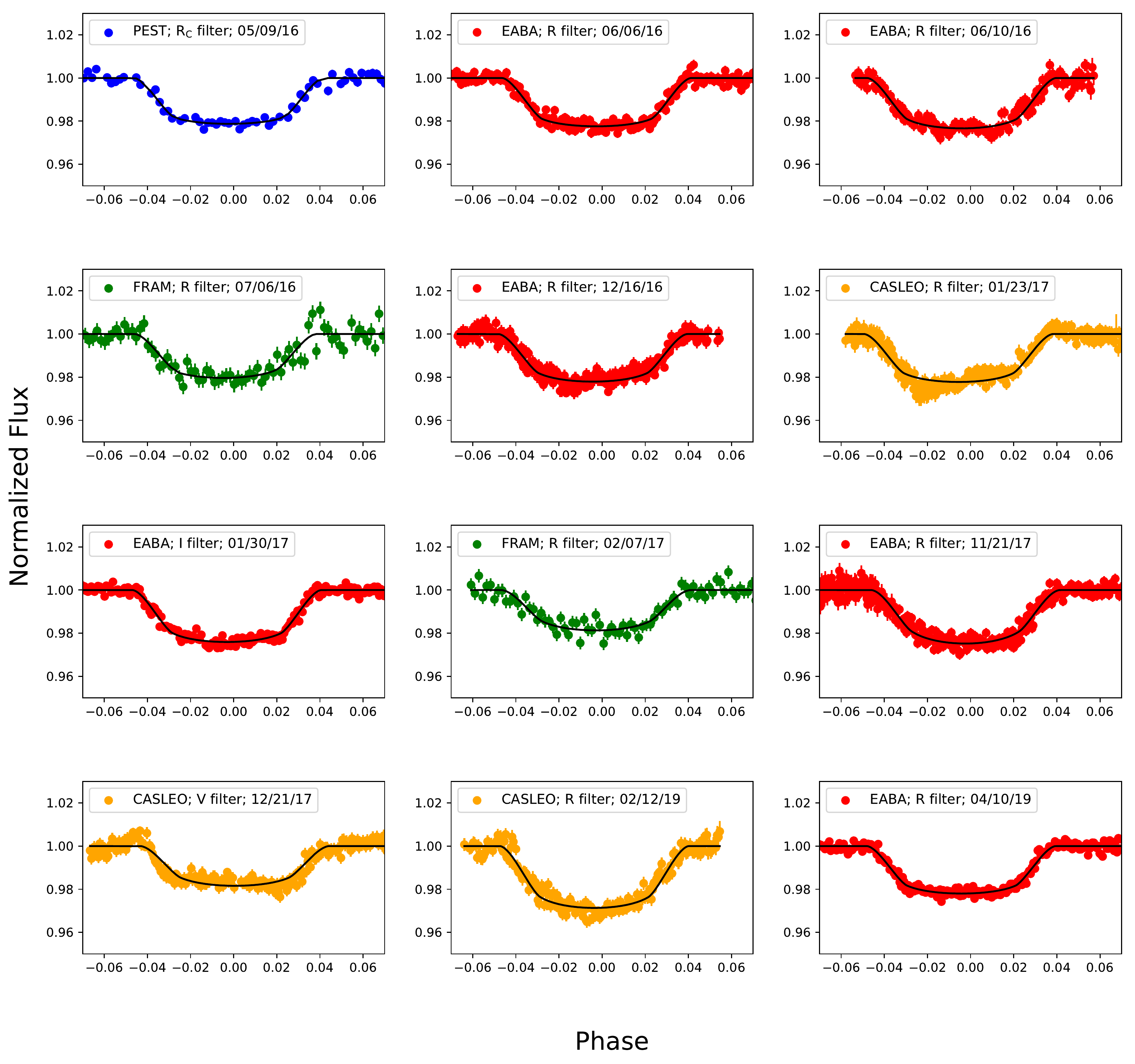}
   \caption{Complete transits of WASP-19b observed by our team. Blue, red, orange, and green colors indicate the photometric data (circles) and error bars of the observations obtained at PEST, EABA, and CASLEO observatories and with the FRAM telescope, respectively. Best-fittings are marked in
solid black lines. For each transit the observatory/telescope, filter, and observation date are also indicated.}
     \label{transitos}
\end{figure*}

We show the new full transits observed by our team and their best fit models in Figure \ref{transitos}, and in Table \ref{table:3}, we list the photometric parameters obtained for the 74 complete transit light curves. 

\begin{table*}
{\scriptsize
\caption{Photometric parameters and quality factors determined for the 74 transit light curves analyzed in this work}          
\label{table:3}      
\centering          
\begin{tabular}{lcccccccccc}     
\hline\hline       
Date	&	Epoch	&	$i$ 	&	$\Sigma$ & $k$ &	 $r_{\star}$ &	$r_\mathrm{P}$ & PNR & $\beta$ & Filter & Reference \\
\hline                   
Dec 17, 2008	&	53	&	80.66	$^{+	1.26	}_{-	1.29	}$ &	0.3046	$^{+	0.0150	}_{-	0.0140	}$ &	0.1391	$^{+	0.0030	}_{-	0.0037	}$ &	0.2674	$^{+	0.0127	}_{-	0.0122	}$ &	0.0483	$^{+	0.0089	}_{-	0.0080	}$ &	0.0013	&	1.2074	&	z	&	1	\\
Feb 28, 2010	&	609	&	78.85	$^{+	0.55	}_{-	0.66	}$ &	0.3282	$^{+	0.0070	}_{-	0.0063	}$ &	0.1433	$^{+	0.0018	}_{-	0.0030	}$ &	0.2872	$^{+	0.0059	}_{-	0.0053	}$ &	0.0374	$^{+	0.0018	}_{-	0.0019	}$ &	0.0009	&	0.9904	&	Gunn-r	&	2	\\
Mar 04, 2010	&	614	&	81.38	$^{+	4.30	}_{-	2.99	}$ &	0.2999	$^{+	0.0336	}_{-	0.0313	}$ &	0.1405	$^{+	0.0048	}_{-	0.0139	}$ &	0.2634	$^{+	0.0285	}_{-	0.0266	}$ &	0.0369	$^{+	0.0049	}_{-	0.0059	}$ &	0.0033	&	0.9773	&	clear	&	3	\\
Mar 18, 2010	&	632	&	80.43	$^{+	5.06	}_{-	4.52	}$ &	0.3257	$^{+	0.0502	}_{-	0.0493	}$ &	0.1416	$^{+	0.0081	}_{-	0.0077	}$ &	0.2859	$^{+	0.0407	}_{-	0.0419	}$ &	0.0415	$^{+	0.0062	}_{-	0.0067	}$ &	0.0054	&	0.8379	&	clear	&	4	\\
May 18, 2010	&	709	&	80.38	$^{+	1.33	}_{-	1.20	}$ &	0.3109	$^{+	0.0119	}_{-	0.0123	}$ &	0.1431	$^{+	0.0028	}_{-	0.0045	}$ &	0.2722	$^{+	0.0101	}_{-	0.0106	}$ &	0.0394	$^{+	0.0066	}_{-	0.0053	}$ &	0.0014	&	0.8135	&	Gunn-i	&	5	\\
May 21, 2010	&	714	&	79.00	$^{+	0.94	}_{-	0.93	}$ &	0.3213	$^{+	0.0107	}_{-	0.0110	}$ &	0.1426	$^{+	0.0016	}_{-	0.0019	}$ &	0.2812	$^{+	0.0092	}_{-	0.0094	}$ &	0.0396	$^{+	0.0014	}_{-	0.0017	}$ &	0.0015	&	0.8101	&	I+z'	&	6	\\
Jun 05, 2010	&	733	&	80.49	$^{+	1.05	}_{-	1.16	}$ &	0.3110	$^{+	0.0105	}_{-	0.0126	}$ &	0.1456	$^{+	0.0020	}_{-	0.0032	}$ &	0.2712	$^{+	0.0090	}_{-	0.0103	}$ &	0.0460	$^{+	0.0020	}_{-	0.0023	}$ &	0.0016	&	0.6083	&	Gunn-i	&	5	\\
Jun 20, 2010	&	752	&	80.25	$^{+	2.05	}_{-	5.02	}$ &	0.3261	$^{+	0.0523	}_{-	0.0185	}$ &	0.1447	$^{+	0.0041	}_{-	0.0120	}$ &	0.2854	$^{+	0.0445	}_{-	0.0159	}$ &	0.0329	$^{+	0.0067	}_{-	0.0064	}$ &	0.003	&	0.714	&	Gunn-i	&	5	\\
Dec 09, 2010	&	969	&	81.14	$^{+	2.41	}_{-	2.03	}$ &	0.3051	$^{+	0.0206	}_{-	0.0243	}$ &	0.1300	$^{+	0.0036	}_{-	0.0068	}$ &	0.2699	$^{+	0.0173	}_{-	0.0168	}$ &	0.0414	$^{+	0.0055	}_{-	0.0062	}$ &	0.0022	&	0.5218	&	IC	&	6	\\
Jan 08, 2011	&	1007	&	79.67	$^{+	1.13	}_{-	1.00	}$ &	0.3228	$^{+	0.0122	}_{-	0.0112	}$ &	0.1416	$^{+	0.0037	}_{-	0.0038	}$ &	0.2824	$^{+	0.0102	}_{-	0.0099	}$ &	0.0386	$^{+	0.0037	}_{-	0.0031	}$ &	0.0015	&	0.97	&	I+z'	&	6	\\
Jan 18, 2011	&	1021	&	80.66	$^{+	2.08	}_{-	2.48	}$ &	0.3020	$^{+	0.0268	}_{-	0.0199	}$ &	0.1387	$^{+	0.0030	}_{-	0.0062	}$ &	0.2653	$^{+	0.0227	}_{-	0.0177	}$ &	0.0370	$^{+	0.0025	}_{-	0.0028	}$ &	0.0024	&	0.9966	&	R	&	7	\\
Jan 23, 2011	&	1026	&	79.87	$^{+	1.19	}_{-	1.06	}$ &	0.3117	$^{+	0.0118	}_{-	0.0129	}$ &	0.1358	$^{+	0.0026	}_{-	0.0028	}$ &	0.2744	$^{+	0.0100	}_{-	0.0113	}$ &	0.0372	$^{+	0.0019	}_{-	0.0020	}$ &	0.0015	&	0.9854	&	I+z'	&	6	\\
Jan 23, 2011	&	1026	&	78.17	$^{+	0.77	}_{-	0.74	}$ &	0.3343	$^{+	0.0076	}_{-	0.0107	}$ &	0.1441	$^{+	0.0020	}_{-	0.0021	}$ &	0.2917	$^{+	0.0070	}_{-	0.0086	}$ &	0.0366	$^{+	0.0077	}_{-	0.0067	}$ &	0.0013	&	0.7031	&	Gunn-r'	&	6	\\
Feb 01, 2011	&	1038	&	83.53	$^{+	6.33	}_{-	3.62	}$ &	0.2759	$^{+	0.0576	}_{-	0.0268	}$ &	0.1365	$^{+	0.0081	}_{-	0.0074	}$ &	0.2424	$^{+	0.0509	}_{-	0.0226	}$ &	0.0361	$^{+	0.0053	}_{-	0.0085	}$ &	0.0058	&	1.2582	&	R$_\mathrm{C}$	&	8	\\
Feb 08, 2011	&	1047	&	77.37	$^{+	3.51	}_{-	2.90	}$ &	0.3501	$^{+	0.0361	}_{-	0.0399	}$ &	0.1471	$^{+	0.0054	}_{-	0.0167	}$ &	0.3064	$^{+	0.0329	}_{-	0.0324	}$ &	0.0454	$^{+	0.0039	}_{-	0.0054	}$ &	0.0037	&	0.9181	&	R$_\mathrm{C}$	&	8	\\
Feb 10, 2011	&	1049	&	77.28	$^{+	1.22	}_{-	1.95	}$ &	0.3482	$^{+	0.0165	}_{-	0.0112	}$ &	0.1518	$^{+	0.0080	}_{-	0.0089	}$ &	0.3017	$^{+	0.0165	}_{-	0.0103	}$ &	0.0463	$^{+	0.0025	}_{-	0.0026	}$ &	0.0017	&	1.0489	&	I+z'	&	6	\\
Feb 14, 2011	&	1054	&	78.57	$^{+	1.27	}_{-	1.43	}$ &	0.3309	$^{+	0.0153	}_{-	0.0146	}$ &	0.1449	$^{+	0.0019	}_{-	0.0035	}$ &	0.2888	$^{+	0.0135	}_{-	0.0127	}$ &	0.0445	$^{+	0.0014	}_{-	0.0029	}$ &	0.0017	&	0.8804	&	Gunn-z'	&	6	\\
Feb 15, 2011	&	1055	&	82.03	$^{+	2.98	}_{-	2.10	}$ &	0.2958	$^{+	0.0217	}_{-	0.0233	}$ &	0.1372	$^{+	0.0047	}_{-	0.0068	}$ &	0.2601	$^{+	0.0180	}_{-	0.0194	}$ &	0.0358	$^{+	0.0036	}_{-	0.0041	}$ &	0.0023	&	0.3502	&	I+z'	&	6	\\
Mar 02, 2011	&	1074	&	79.86	$^{+	2.03	}_{-	1.48	}$ &	0.3190	$^{+	0.0146	}_{-	0.0198	}$ &	0.1407	$^{+	0.0041	}_{-	0.0053	}$ &	0.2797	$^{+	0.0124	}_{-	0.0163	}$ &	0.0370	$^{+	0.0075	}_{-	0.0061	}$ &	0.0019	&	0.426	&	I+z'	&	6	\\
Mar 03, 2011	&	1076	&	79.43	$^{+	6.75	}_{-	5.21	}$ &	0.3230	$^{+	0.0698	}_{-	0.0415	}$ &	0.1329	$^{+	0.0082	}_{-	0.0167	}$ &	0.2838	$^{+	0.0634	}_{-	0.0354	}$ &	0.0393	$^{+	0.0057	}_{-	0.0086	}$ &	0.0048	&	0.937	&	R$_\mathrm{C}$	&	8	\\
Apr 04, 2011	&	1116	&	81.47	$^{+	1.47	}_{-	0.94	}$ &	0.2911	$^{+	0.0110	}_{-	0.0173	}$ &	0.1354	$^{+	0.0038	}_{-	0.0044	}$ &	0.2562	$^{+	0.0093	}_{-	0.0145	}$ &	0.0399	$^{+	0.0027	}_{-	0.0023	}$ &	0.0017	&	1.6274	&	I+z'	&	6	\\
May 08, 2011	&	1159	&	80.01	$^{+	1.68	}_{-	2.02	}$ &	0.3161	$^{+	0.0227	}_{-	0.0216	}$ &	0.1395	$^{+	0.0047	}_{-	0.0059	}$ &	0.2773	$^{+	0.0193	}_{-	0.0191	}$ &	0.0380	$^{+	0.0028	}_{-	0.0020	}$ &	0.0028	&	1.1419	&	Gunn-i	&	5	\\
May 11, 2011	&	1164	&	79.52	$^{+	0.95	}_{-	1.38	}$ &	0.3204	$^{+	0.0355	}_{-	0.0118	}$ &	0.1402	$^{+	0.0076	}_{-	0.0138	}$ &	0.2803	$^{+	0.0342	}_{-	0.0101	}$ &	0.0395	$^{+	0.0016	}_{-	0.0019	}$ &	0.0012	&	1.217	&	Gunn-i	&	5	\\
May 22, 2011	&	1178	&	77.50	$^{+	2.22	}_{-	1.67	}$ &	0.3400	$^{+	0.0174	}_{-	0.0281	}$ &	0.1367	$^{+	0.0027	}_{-	0.0060	}$ &	0.2985	$^{+	0.0149	}_{-	0.0239	}$ &	0.0369	$^{+	0.0042	}_{-	0.0061	}$ &	0.0025	&	0.6181	&	Gunn-i	&	5	\\
May 26, 2011	&	1183	&	81.50	$^{+	1.63	}_{-	1.42	}$ &	0.3031	$^{+	0.0286	}_{-	0.0128	}$ &	0.1370	$^{+	0.0105	}_{-	0.0081	}$ &	0.2663	$^{+	0.0266	}_{-	0.0122	}$ &	0.0354	$^{+	0.0033	}_{-	0.0030	}$ &	0.0015	&	0.7159	&	Gunn-i	&	5	\\
Mar 13, 2012	&	1552	&	78.66	$^{+	0.96	}_{-	1.18	}$ &	0.3274	$^{+	0.0142	}_{-	0.0114	}$ &	0.1407	$^{+	0.0014	}_{-	0.0024	}$ &	0.2869	$^{+	0.0132	}_{-	0.0096	}$ &	0.0387	$^{+	0.0063	}_{-	0.0091	}$ &	0.0017	&	1.0022	&	1.25-1.33 $\mu$m	&	9	\\
Mar 13, 2012	&	1552	&	79.11	$^{+	1.26	}_{-	1.00	}$ &	0.3282	$^{+	0.0120	}_{-	0.0139	}$ &	0.1450	$^{+	0.0025	}_{-	0.0032	}$ &	0.2856	$^{+	0.0113	}_{-	0.0116	}$ &	0.0384	$^{+	0.0039	}_{-	0.0050	}$ &	0.0018	&	1.2523	&	1.4-1.5 $\mu$m	&	9	\\
Mar 13, 2012	&	1552	&	79.70	$^{+	0.56	}_{-	0.52	}$ &	0.3196	$^{+	0.0038	}_{-	0.0064	}$ &	0.1564	$^{+	0.0012	}_{-	0.0022	}$ &	0.2764	$^{+	0.0035	}_{-	0.0057	}$ &	0.0431	$^{+	0.0012	}_{-	0.0014	}$ &	0.0012	&	0.8852	&	1.5-1.6 $\mu$m	&	9	\\
Mar 13, 2012	&	1552	&	78.82	$^{+	0.83	}_{-	0.68	}$ &	0.3271	$^{+	0.0079	}_{-	0.0095	}$ &	0.1434	$^{+	0.0022	}_{-	0.0025	}$ &	0.2860	$^{+	0.0071	}_{-	0.0084	}$ &	0.0322	$^{+	0.0079	}_{-	0.0038	}$ &	0.0012	&	0.9867	&	1.6-1.7 $\mu$m	&	9	\\
Mar 13, 2012	&	1552	&	79.40	$^{+	0.71	}_{-	0.74	}$ &	0.3207	$^{+	0.0103	}_{-	0.0101	}$ &	0.1376	$^{+	0.0029	}_{-	0.0031	}$ &	0.2820	$^{+	0.0086	}_{-	0.0089	}$ &	0.0414	$^{+	0.0040	}_{-	0.0076	}$ &	0.0014	&	1.2102	&	1.7-1.8 $\mu$m	&	9	\\
Mar 13, 2012	&	1552	&	78.90	$^{+	0.91	}_{-	0.91	}$ &	0.3246	$^{+	0.0106	}_{-	0.0105	}$ &	0.1436	$^{+	0.0015	}_{-	0.0016	}$ &	0.2838	$^{+	0.0093	}_{-	0.0090	}$ &	0.0391	$^{+	0.0016	}_{-	0.0017	}$ &	0.0018	&	0.571	&	1.95-2.05 $\mu$m	&	9	\\
Mar 13, 2012	&	1552	&	77.93	$^{+	0.69	}_{-	0.60	}$ &	0.3377	$^{+	0.0070	}_{-	0.0075	}$ &	0.1441	$^{+	0.0011	}_{-	0.0011	}$ &	0.2948	$^{+	0.0066	}_{-	0.0063	}$ &	0.0345	$^{+	0.0022	}_{-	0.0024	}$ &	0.0012	&	0.7515	&	2.05-2.15 $\mu$m	&	9	\\
Mar 13, 2012	&	1552	&	79.95	$^{+	0.96	}_{-	0.80	}$ &	0.3123	$^{+	0.0088	}_{-	0.0105	}$ &	0.1413	$^{+	0.0012	}_{-	0.0024	}$ &	0.2737	$^{+	0.0074	}_{-	0.0094	}$ &	0.0349	$^{+	0.0157	}_{-	0.0056	}$ &	0.0015	&	0.815	&	2.15-2.25 $\mu$m	&	9	\\
Mar 13, 2012	&	1552	&	82.07	$^{+	2.53	}_{-	2.24	}$ &	0.2917	$^{+	0.0242	}_{-	0.0203	}$ &	0.1375	$^{+	0.0030	}_{-	0.0041	}$ &	0.2565	$^{+	0.0208	}_{-	0.0170	}$ &	0.0351	$^{+	0.0030	}_{-	0.0048	}$ &	0.0025	&	0.8573	&	2.25-2.35 $\mu$m	&	9	\\
Apr 04, 2012	&	1580	&	76.70	$^{+	0.79	}_{-	0.68	}$ &	0.3532	$^{+	0.0073	}_{-	0.0093	}$ &	0.1484	$^{+	0.0014	}_{-	0.0018	}$ &	0.3076	$^{+	0.0069	}_{-	0.0083	}$ &	0.0371	$^{+	0.0027	}_{-	0.0022	}$ &	0.0014	&	0.8314	&	1.25-1.33 $\mu$m	&	9	\\
Apr 04, 2012	&	1580	&	78.28	$^{+	0.82	}_{-	0.79	}$ &	0.3347	$^{+	0.0093	}_{-	0.0093	}$ &	0.1473	$^{+	0.0013	}_{-	0.0014	}$ &	0.2916	$^{+	0.0081	}_{-	0.0079	}$ &	0.0388	$^{+	0.0037	}_{-	0.0058	}$ &	0.0016	&	0.7089	&	1.4-1.5 $\mu$m	&	9	\\
Apr 04, 2012	&	1580	&	78.59	$^{+	0.70	}_{-	1.07	}$ &	0.3320	$^{+	0.0121	}_{-	0.0075	}$ &	0.1461	$^{+	0.0013	}_{-	0.0016	}$ &	0.2897	$^{+	0.0103	}_{-	0.0064	}$ &	0.0418	$^{+	0.0009	}_{-	0.0013	}$ &	0.0011	&	0.7849	&	1.5-1.6 $\mu$m	&	9	\\
Apr 04, 2012	&	1580	&	78.77	$^{+	0.70	}_{-	0.60	}$ &	0.3293	$^{+	0.0067	}_{-	0.0069	}$ &	0.1453	$^{+	0.0013	}_{-	0.0025	}$ &	0.2875	$^{+	0.0062	}_{-	0.0063	}$ &	0.0427	$^{+	0.0100	}_{-	0.0077	}$ &	0.0011	&	1.0223	&	1.6-1.7 $\mu$m	&	9	\\
Apr 04, 2012	&	1580	&	78.04	$^{+	0.54	}_{-	0.63	}$ &	0.3379	$^{+	0.0068	}_{-	0.0069	}$ &	0.1437	$^{+	0.0012	}_{-	0.0014	}$ &	0.2954	$^{+	0.0060	}_{-	0.0060	}$ &	0.0395	$^{+	0.0018	}_{-	0.0018	}$ &	0.0012	&	0.8126	&	1.7-1.8 $\mu$m	&	9	\\
Apr 04, 2012	&	1580	&	79.08	$^{+	0.86	}_{-	0.68	}$ &	0.3279	$^{+	0.0080	}_{-	0.0102	}$ &	0.1513	$^{+	0.0012	}_{-	0.0019	}$ &	0.2846	$^{+	0.0073	}_{-	0.0087	}$ &	0.0385	$^{+	0.0038	}_{-	0.0027	}$ &	0.0017	&	0.9655	&	1.95-2.05 $\mu$m	&	9	\\
Apr 04, 2012	&	1580	&	77.93	$^{+	0.70	}_{-	0.60	}$ &	0.3353	$^{+	0.0077	}_{-	0.0074	}$ &	0.1441	$^{+	0.0014	}_{-	0.0017	}$ &	0.2930	$^{+	0.0068	}_{-	0.0071	}$ &	0.0403	$^{+	0.0021	}_{-	0.0023	}$ &	0.0011	&	0.9173	&	2.05-2.15 $\mu$m	&	9	\\
Apr 04, 2012	&	1580	&	79.29	$^{+	1.09	}_{-	0.80	}$ &	0.3217	$^{+	0.0089	}_{-	0.0122	}$ &	0.1406	$^{+	0.0011	}_{-	0.0017	}$ &	0.2821	$^{+	0.0077	}_{-	0.0106	}$ &	0.0374	$^{+	0.0044	}_{-	0.0045	}$ &	0.0014	&	0.6433	&	2.15-2.25 $\mu$m	&	9	\\
Apr 04, 2012	&	1580	&	79.51	$^{+	1.34	}_{-	1.01	}$ &	0.3196	$^{+	0.0137	}_{-	0.0159	}$ &	0.1434	$^{+	0.0045	}_{-	0.0040	}$ &	0.2799	$^{+	0.0112	}_{-	0.0138	}$ &	0.0399	$^{+	0.0018	}_{-	0.0019	}$ &	0.0023	&	1.262	&	2.25-2.35 $\mu$m	&	9	\\
Apr 15, 2012	&	1595	&	78.23	$^{+	1.23	}_{-	1.56	}$ &	0.3365	$^{+	0.0201	}_{-	0.0175	}$ &	0.1455	$^{+	0.0048	}_{-	0.0039	}$ &	0.2936	$^{+	0.0170	}_{-	0.0149	}$ &	0.0406	$^{+	0.0090	}_{-	0.0079	}$ &	0.0019	&	0.9515	&	H	&	5	\\
Apr 15, 2012	&	1595	&	80.98	$^{+	1.58	}_{-	1.21	}$ &	0.3018	$^{+	0.0143	}_{-	0.0167	}$ &	0.1320	$^{+	0.0026	}_{-	0.0029	}$ &	0.2669	$^{+	0.0131	}_{-	0.0153	}$ &	0.0390	$^{+	0.0021	}_{-	0.0020	}$ &	0.0014	&	1.1579	&	J	&	5	\\
Apr 15, 2012	&	1595	&	82.51	$^{+	3.05	}_{-	2.50	}$ &	0.2837	$^{+	0.0263	}_{-	0.0253	}$ &	0.1332	$^{+	0.0051	}_{-	0.0034	}$ &	0.2506	$^{+	0.0232	}_{-	0.0230	}$ &	0.0398	$^{+	0.0021	}_{-	0.0022	}$ &	0.0022	&	1.4577	&	K	&	5	\\
May 15, 2012	&	1633	&	76.32	$^{+	1.60	}_{-	1.94	}$ &	0.3558	$^{+	0.0229	}_{-	0.0218	}$ &	0.1432	$^{+	0.0021	}_{-	0.0084	}$ &	0.3108	$^{+	0.0220	}_{-	0.0180	}$ &	0.0455	$^{+	0.0072	}_{-	0.0065	}$ &	0.0019	&	0.746	&	I+z'	&	6	\\
Feb 01, 2013	&	1964	&	77.03	$^{+	4.55	}_{-	3.03	}$ &	0.3526	$^{+	0.0350	}_{-	0.0438	}$ &	0.1352	$^{+	0.0066	}_{-	0.0173	}$ &	0.3101	$^{+	0.0325	}_{-	0.0368	}$ &	0.0445	$^{+	0.0051	}_{-	0.0081	}$ &	0.0051	&	0.6298	&	clear	&	10	\\
Feb 11, 2013	&	1977	&	79.08	$^{+	2.98	}_{-	3.00	}$ &	0.3203	$^{+	0.0377	}_{-	0.0295	}$ &	0.1408	$^{+	0.0107	}_{-	0.0098	}$ &	0.2804	$^{+	0.0341	}_{-	0.0239	}$ &	0.0416	$^{+	0.0052	}_{-	0.0056	}$ &	0.0045	&	0.771	&	clear	&	4	\\
Apr 08, 2013	&	2048	&	80.26	$^{+	2.05	}_{-	2.61	}$ &	0.3110	$^{+	0.0327	}_{-	0.0282	}$ &	0.1365	$^{+	0.0064	}_{-	0.0097	}$ &	0.2731	$^{+	0.0276	}_{-	0.0236	}$ &	0.0386	$^{+	0.0046	}_{-	0.0047	}$ &	0.0033	&	0.113	&	clear	&	4	\\
May 24, 2013	&	2106	&	81.65	$^{+	7.14	}_{-	2.96	}$ &	0.3002	$^{+	0.0353	}_{-	0.0453	}$ &	0.1419	$^{+	0.0094	}_{-	0.0081	}$ &	0.2627	$^{+	0.0280	}_{-	0.0382	}$ &	0.0370	$^{+	0.0074	}_{-	0.0065	}$ &	0.0072	&	0.7942	&	clear	&	10	\\
Mar 03, 2014	&	2465	&	82.01	$^{+	7.79	}_{-	4.98	}$ &	0.2884	$^{+	0.0698	}_{-	0.0325	}$ &	0.1287	$^{+	0.0071	}_{-	0.0179	}$ &	0.2565	$^{+	0.0599	}_{-	0.0288	}$ &	0.0328	$^{+	0.0068	}_{-	0.0064	}$ &	0.0044	&	0.7579	&	clear	&	4	\\
Mar 22, 2014	&	2490	&	79.24	$^{+	0.94	}_{-	1.76	}$ &	0.3207	$^{+	0.0239	}_{-	0.0159	}$ &	0.1361	$^{+	0.0045	}_{-	0.0060	}$ &	0.2823	$^{+	0.0189	}_{-	0.0130	}$ &	0.0387	$^{+	0.0014	}_{-	0.0015	}$ &	0.0009	&	0.8471	&	clear	&	11	\\
Mar 23, 2014	&	2490	&	77.99	$^{+	2.27	}_{-	1.17	}$ &	0.3364	$^{+	0.0127	}_{-	0.0253	}$ &	0.1425	$^{+	0.0026	}_{-	0.0104	}$ &	0.2945	$^{+	0.0105	}_{-	0.0219	}$ &	0.0417	$^{+	0.0030	}_{-	0.0041	}$ &	0.0025	&	0.6102	&	clear	&	12	\\
Nov 15, 2014	&	2792	&	80.06	$^{+	0.19	}_{-	0.20	}$ &	0.3176	$^{+	0.0021	}_{-	0.0022	}$ &	0.1390	$^{+	0.0014	}_{-	0.0016	}$ &	0.2787	$^{+	0.0020	}_{-	0.0019	}$ &	0.0417	$^{+	0.0015	}_{-	0.0018	}$ &	0.0002	&	1.3064	&	clear	&	13	\\
Jan 30, 2016	&	3351	&	78.33	$^{+	0.47	}_{-	0.36	}$ &	0.3324	$^{+	0.0048	}_{-	0.0052	}$ &	0.1392	$^{+	0.0018	}_{-	0.0022	}$ &	0.2919	$^{+	0.0042	}_{-	0.0046	}$ &	0.0410	$^{+	0.0011	}_{-	0.0014	}$ &	0.0005	&	2.089	&	clear	&	13	\\
Feb 29, 2016	&	3389	&	79.46	$^{+	0.31	}_{-	0.38	}$ &	0.3205	$^{+	0.0038	}_{-	0.0030	}$ &	0.1474	$^{+	0.0017	}_{-	0.0018	}$ &	0.2793	$^{+	0.0035	}_{-	0.0028	}$ &	0.0428	$^{+	0.0035	}_{-	0.0030	}$ &	0.0005	&	0.5848	&	clear	&	13	\\
May 09, 2016	&	3477	&	80.03	$^{+	3.83	}_{-	2.99	}$ &	0.3167	$^{+	0.0362	}_{-	0.0322	}$ &	0.1391	$^{+	0.0041	}_{-	0.0107	}$ &	0.2782	$^{+	0.0318	}_{-	0.0274	}$ &	0.0387	$^{+	0.0044	}_{-	0.0058	}$ &	0.0034	&	0.8907	&	R$_\mathrm{C}$	&	14	\\
Jun 06, 2016	&	3513	&	79.39	$^{+	2.13	}_{-	2.66	}$ &	0.3208	$^{+	0.0297	}_{-	0.0216	}$ &	0.1442	$^{+	0.0028	}_{-	0.0096	}$ &	0.2800	$^{+	0.0263	}_{-	0.0185	}$ &	0.0409	$^{+	0.0022	}_{-	0.0025	}$ &	0.0021	&	1.1813	&	R	&	15	\\
Jun 10, 2016	&	3518	&	79.20	$^{+	2.87	}_{-	2.30	}$ &	0.3234	$^{+	0.0334	}_{-	0.0378	}$ &	0.1472	$^{+	0.0075	}_{-	0.0064	}$ &	0.2813	$^{+	0.0275	}_{-	0.0312	}$ &	0.0459	$^{+	0.0107	}_{-	0.0052	}$ &	0.0029	&	1.0471	&	R	&	15	\\
Jul 06, 2016	&	3551	&	80.52	$^{+	5.97	}_{-	3.58	}$ &	0.3130	$^{+	0.0467	}_{-	0.0506	}$ &	0.1356	$^{+	0.0094	}_{-	0.0086	}$ &	0.2755	$^{+	0.0392	}_{-	0.0430	}$ &	0.0374	$^{+	0.0079	}_{-	0.0074	}$ &	0.006	&	0.8156	&	R	&	16	\\
Dec 02, 2016	&	3739	&	78.86	$^{+	1.19	}_{-	1.02	}$ &	0.3300	$^{+	0.0242	}_{-	0.0173	}$ &	0.1616	$^{+	0.0299	}_{-	0.0196	}$ &	0.2848	$^{+	0.0126	}_{-	0.0152	}$ &	0.0299	$^{+	0.0072	}_{-	0.0024	}$ &	0.0015	&	2.4622	&	I	&	17	\\
Dec 16, 2016	&	3758	&	77.92	$^{+	2.64	}_{-	1.81	}$ &	0.3465	$^{+	0.0250	}_{-	0.0352	}$ &	0.1521	$^{+	0.0144	}_{-	0.0130	}$ &	0.2998	$^{+	0.0206	}_{-	0.0286	}$ &	0.0462	$^{+	0.0030	}_{-	0.0038	}$ &	0.0019	&	1.7937	&	R	&	15	\\
\hline
\end{tabular}}
\end{table*}

\begin{table*}
\renewcommand\thetable{3}
{\scriptsize
\caption{Continued}            
\label{table:3}      
\centering          
\begin{tabular}{lcccccccccc}     
\hline\hline       
Date	&	Epoch	&	$i$ 	&	$\Sigma$ & $k$ &	 $r_{\star}$ &	$r_\mathrm{P}$ & PNR & $\beta$ & Filter & Reference \\
\hline                   
Jan 23, 2017	&	3806	&	80.24	$^{+	2.38	}_{-	3.22	}$ &	0.3077	$^{+	0.0405	}_{-	0.0316	}$ &	0.1436	$^{+	0.0080	}_{-	0.0070	}$ &	0.2686	$^{+	0.0346	}_{-	0.0264	}$ &	0.0345	$^{+	0.0042	}_{-	0.0048	}$ &	0.0024	&	2.8005	&	R	&	18	\\
Jan 30, 2017	&	3815	&	79.93	$^{+	2.45	}_{-	1.49	}$ &	0.3195	$^{+	0.0173	}_{-	0.0205	}$ &	0.1415	$^{+	0.0035	}_{-	0.0091	}$ &	0.2795	$^{+	0.0149	}_{-	0.0170	}$ &	0.0395	$^{+	0.0024	}_{-	0.0029	}$ &	0.0018	&	1.1944	&	I	&	15	\\
Feb 07, 2017	&	3825	&	74.51	$^{+	3.79	}_{-	2.27	}$ &	0.3841	$^{+	0.0336	}_{-	0.0430	}$ &	0.1322	$^{+	0.0094	}_{-	0.0185	}$ &	0.3314	$^{+	0.0376	}_{-	0.0286	}$ &	0.0445	$^{+	0.0060	}_{-	0.0066	}$ &	0.0048	&	0.6168	&	R	&	16	\\
Feb 11, 2017	&	3830	&	80.68	$^{+	3.09	}_{-	2.36	}$ &	0.2977	$^{+	0.0293	}_{-	0.0241	}$ &	0.1295	$^{+	0.0044	}_{-	0.0084	}$ &	0.2642	$^{+	0.0236	}_{-	0.0206	}$ &	0.0368	$^{+	0.0038	}_{-	0.0047	}$ &	0.0009	&	2.3335	&	clear	&	12	\\
Apr 12, 2017	&	3906	&	78.78	$^{+	0.90	}_{-	0.76	}$ &	0.3254	$^{+	0.0108	}_{-	0.0120	}$ &	0.1475	$^{+	0.0034	}_{-	0.0046	}$ &	0.2834	$^{+	0.0099	}_{-	0.0101	}$ &	0.0379	$^{+	0.0028	}_{-	0.0027	}$ &	0.0011	&	2.3205	&	clear	&	12	\\
Nov 21, 2017	&	4189	&	77.73	$^{+	2.03	}_{-	1.54	}$ &	0.3459	$^{+	0.0160	}_{-	0.0208	}$ &	0.1563	$^{+	0.0042	}_{-	0.0122	}$ &	0.2990	$^{+	0.0139	}_{-	0.0165	}$ &	0.0465	$^{+	0.0023	}_{-	0.0028	}$ &	0.0021	&	0.984	&	R	&	15	\\
Dec 21, 2017	&	4227	&	82.76	$^{+	7.10	}_{-	6.05	}$ &	0.3080	$^{+	0.0867	}_{-	0.0397	}$ &	0.1361	$^{+	0.0209	}_{-	0.0115	}$ &	0.2692	$^{+	0.0706	}_{-	0.0314	}$ &	0.0354	$^{+	0.0041	}_{-	0.0037	}$ &	0.0025	&	2.0785	&	V	&	18	\\
Apr 04, 2018	&	4359	&	79.54	$^{+	3.06	}_{-	3.00	}$ &	0.3118	$^{+	0.0376	}_{-	0.0323	}$ &	0.1393	$^{+	0.0024	}_{-	0.0066	}$ &	0.2733	$^{+	0.0327	}_{-	0.0284	}$ &	0.0390	$^{+	0.0016	}_{-	0.0017	}$ &	0.0017	&	1.2029	&	R	&	19	\\
Feb 12, 2019	&	4757	&	77.97	$^{+	1.81	}_{-	1.59	}$ &	0.3416	$^{+	0.0300	}_{-	0.0357	}$ &	0.1664	$^{+	0.0186	}_{-	0.0141	}$ &	0.2926	$^{+	0.0206	}_{-	0.0277	}$ &	0.0420	$^{+	0.0019	}_{-	0.0042	}$ &	0.0027	&	0.8879	&	R	&	18	\\
Mar 06, 2019	&	4784	&	83.64	$^{+	6.26	}_{-	3.93	}$ &	0.2712	$^{+	0.0651	}_{-	0.0235	}$ &	0.1311	$^{+	0.0076	}_{-	0.0056	}$ &	0.2402	$^{+	0.0545	}_{-	0.0202	}$ &	0.0320	$^{+	0.0067	}_{-	0.0042	}$ &	0.0044	&	0.7301	&	V	&	20	\\
Apr 10, 2019	&	4829	&	79.84	$^{+	1.07	}_{-	0.93	}$ &	0.3117	$^{+	0.0115	}_{-	0.0113	}$ &	0.1436	$^{+	0.0021	}_{-	0.0029	}$ &	0.2724	$^{+	0.0100	}_{-	0.0095	}$ &	0.0406	$^{+	0.0030	}_{-	0.0035	}$ &	0.0015	&	0.8415	&	R	&	15	\\
\hline      
\end{tabular}}

Columns 3-7: Values of the derived photometric parameters and their errors. Column 8: Photometric noise rate. Column 9: Median value for the red noise. 

References: (1) \citet{hebb}; (2) \citet{hellier}; (3) Colque J. (TRESCA); (4) Evans P. (TRESCA); (5) \citet{mancini}; (6) \citet{lendl}; (7) \citet{dragomir}; (8) T.G. Tan (TRESCA); (9) \citet{bean}; (10) Ma\v{s}ek M.; (11) \citet{espinoza}; (12) Ma\v{s}ek M., Ho\v{n}kov\'a K., Jury\v{s}ek J. (TRESCA); (13) \citet{seda}; (14) This work (PEST); (15) This work (EABA); (16) This work (FRAM); (17) Eduardo Fern\'andez-Laj\'us, Romina P. Di Sisto (TRESCA); (18) This work (CASLEO); (19) Carl R Knight (TRESCA); (20) Ana\"{e}l W\"{u}nsche (TRESCA).
\end{table*}

\section{ASSESSMENT OF THE SYSTEMATIC UNCERTAINTIES ON THE MID-TRANSIT TIME MEASUREMENTS}\label{section:4}

As can be noticed in columns 8 and 9 of Table \ref{table:3}, our sample is composed of different quality transits, with a wide range of values of PNR and levels of red noise. The higher values of $\beta$ imply the presence of systematics in the light curve which may have a non-astrophysical origin (such as that related with changes during the observations in the environmental and atmospheric variables i.e., temperature, airmass, fwhm, level of counts from the sky, etc, also introduced by the instruments used to acquire the data, shifts in the position of the star on the CCD, bad correction for pixel sensitivity, etc), or having an astrophysical nature (such as the systematics produced by stellar activity). This can be the case for WASP-19 which is an active star and might present a modulation or anomalies in the light curves due to spot-crossing events \citep{hebb, tregloan}. 
Therefore, given that the main purpose of this work is to evaluate the existence of orbital decay, which depends directly on the errors in the mid-transit times measurements, we explored, in the first place, the influence of the values of the photometric parameters on the mid-times and, afterwards, the influence of the data quality and symmetry of the light curve. 

\begin{figure}
   \centering
   \includegraphics[width=.4\textwidth]{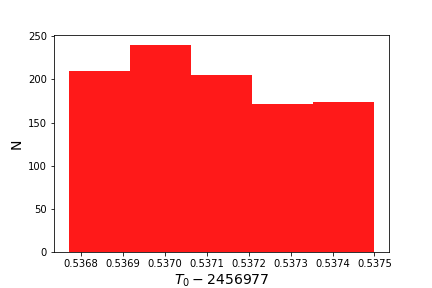}
   \includegraphics[width=.4\textwidth]{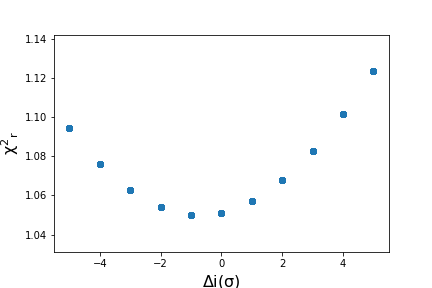}
      \caption{Results for the transit observed on April 15, 2012 in the K-band. \textit{Top panel:} Distribution of mid-transit times obtained after 1000 runs of JKTEBOP considering the adopted values of all the photometric parameters freely varying. \textit{Bottom panel:} Reduced chi-square values of the models obtained by varying the adopted value of the inclination in $-$5 to $+$5 times the error, $\sigma$, computed by \citet{mancini}.}
         \label{figure:2}
\end{figure}

With the first goal in mind, we carried out the three steps procedure described below: 

\begin{itemize}

\item
First, we assessed how different the values of the photometric parameters can be from those of \citet{mancini}, in order to still have a model that properly describes the observations. To do this, we ran JKTEBOP by keeping $i$, $\Sigma$, and $k$ fixed, and sampling $\pm 5 \sigma$ around those estimated by \citet{mancini}. The photometric parameters $l_{0}$ and $T_\mathrm{0}$ were fitted, while $q_{1}$ and $q_{2}$ were allowed to freely varying or not depending on the option chosen in Section \ref{section:3} about how to manage the LD coefficients. As values of these four parameters, we assumed the same considered in Section \ref{section:3}.\\ 

\item
Then, after visually inspecting the observations and their models, we defined for each transit ranges for the values of the photometric parameters in which it is warranted that the model computed with JKTEBOP, correctly represents the observations. In all these cases, we found that the $\Delta\chi^{2} = \chi^{2}_\mathrm{mod}-\chi^{2}_\mathrm{min}$ of the model is well within $1 \sigma$ i.e., the 68.3$\%$ confidence level. We performed this step to avoid adopting values for the photometric parameters that lead to models that do not properly describe the data.\\   

\item
We performed 1000 runs of JKTEBOP randomly varying the adopted values of the photometric parameters: $i$, $\Sigma$, and $k$ (all together and also individually\footnote{In this case, when the adopted value of only one of the parameters was varied, the other two were kept fixed to those obtained by \citet{mancini}.}) in the ranges defined in the previous step. For each transit light curve, we computed the mean value and standard deviation of the mid-transit times given as outputs of these different runs. In Figure \ref{figure:2}, we show the distribution of mid-transit times for the transit observed on April 15, 2012 in the K-band, where the adopted values of all the photometric parameters were allowed to vary (top panel), and the reduced chi-square (between 1.05 and 1.12) resulting after varying the adopted value of the inclination within $\pm 5\sigma$ of measurements by \citealt{mancini} (bottom panel).  

\end{itemize}

One important point to notice here is that the inclination is the photometric parameter that most strongly affects the measured mid-transit times, producing variations as large as 20 seconds in the most extreme case. The values of depth and sum of relative radii play a minor role, causing variations in the mid-transit times of up to 1 second and less than 1 second, respectively.
Additionally, as mentioned in Section \ref{section:3}, the influence of the values of the LD coefficients on the error in the measurement of $T_\mathrm{0}$ is negligible.

Furthermore, we checked if the models for the mid-times found by JKTEBOP by changing the adopted values of the photometric parameters are related to the quality of the light curve (level of white/red noise). Then, after applying the procedure described above, first we searched for possible correlations between the red noise level and standard deviation found for the mid-transit time ($\sigma_{\mathrm{T_0}}$) of each light curve. In order to remove any possible influence of different values of PNR on those of $\sigma_{\mathrm{T_0}}$, we analyzed the transits in three distinct groups depending on their measured photometric noise rate: those with PNR $\leq$ 1.5 mmag (named ``Group 1''), the ones with 1.5 mmag $<$ PNR $\leq$ 3 mmag (Group 2) and finally, transits with PNR $>$ 3 mmag (Group 3). We also considered a group consisting of all the 74 transits (``Group All-$\beta$"). For each of these datasets, we computed the Pearson coefficient\footnote{In order to decide if a positive or negative correlation is weak, moderate or strong we adopted the following criteria, already used in \citet{petrucci18}: i) strongly correlated parameters if -1 $\leq$ r $\lesssim$ -0.8 or 0.8 $\lesssim$ r $\leq$ 1, ii) moderatly correlated parameters if -0.8 $\lesssim$ r $\lesssim$ -0.5 or 0.5 $\lesssim$ r $\lesssim$ 0.8, and weakly correlated parameters if -0.5 $\lesssim$ r $\leq$ 0 or 0 $\leq$ r $\lesssim$ 0.5.} ($r$) and the p-value between $\beta$ and $\sigma_\mathrm{T_0}$. These results are summarized in Table \ref{table:4}. In all the four cases, it is possible to see that $r$ $<$ 0.52 indicating weakly correlated parameters.\\

\begin{table}
\renewcommand\thetable{4}
\caption{Correlation between $\beta$ and $\sigma_{\mathrm{T_0}}$}             
\label{table:4}      
\centering          
\begin{tabular}{ccccc}     
\hline\hline       
 & Group 1 & Group 2 & Group 3 & Group All-$\beta$ \\
\hline                    
$r$	& 0.224 & 0.515 & 0.035  & 0.262 \\
p-value	& 0.241 & 0.003	& 0.900 & 0.023 \\
\hline                  
\end{tabular}
\end{table}

We also explored possible correlations between the values of PNR and $\sigma_{\mathrm{T_0}}$ for each light curve. In this case, in order to eliminate the effect of $\beta$ on the values of  $\sigma_\mathrm{T_0}$, we separated the light curves in three different groups depending on their measured red noise level: those with $\beta < 1$ (Group A), the ones with 1 $\leq$ $\beta$ $<$ 2 (Group B), and transits with $\beta$ $\geq$ 2 (Group C). Same as before, we compared the results including all 74 transits in ``Group All-PNR". In Table \ref{table:5}, we show the Pearson coefficient and p-value between PNR and $\sigma_\mathrm{T_0}$ calculated for each group. Here, it can be seen that for groups A, B and ``All-PNR" the parameters are weakly correlated ($r$ $\leq$ 0.5), but for Group C, $r$ $=$ 0.769 points out a moderate correlation between the parameters. However, we caution that this correlation is based only on 6 points and the p-value is quite high, making this trend not very reliable.

\begin{table}
\renewcommand\thetable{5}
\caption{Correlation between PNR and $\sigma_{\mathrm{T_0}}$}             
\label{table:5}      
\centering          
\begin{tabular}{ccccc}     
\hline\hline       
 & Group A & Group B & Group C & Group All-PNR\\
\hline                    
$r$	&  0.378 &  0.140 & 0.769  & 0.248 \\
p-value	& 0.007 & 0.565 & 0.073 & 0.032 \\
\hline                  
\end{tabular}
\end{table}

Additionally, transit light curves with a high degree of systematics and large dispersion in their photometric data-points tend to have their 4 contact points poorly defined. This makes it difficult for the model to correctly identify the ingress and egress of the transit, which causes a poorly constrained value for the inclination of the system and hence a mid-transit time that is not well constrained. Having this in mind, for all the transits in our sample, we measured the values of $\beta$ and PNR in sections of the light curve related to the estimation of the inclination. These sections were defined: i) Along the complete extension of the transit, ii) Before contact point 1, iii) Between contact points 1 and 2 (ingress), iv) Between contact points 2 and 3 (flat bottom), v) Between contact points 3 and 4 (egress), vi) After contact point 4, vii) Before contact point 1 and after contact point 4 (i.e., the out-of-transit), and viii) Between contact points 1 and 2 and contact points 3 and 4 (i.e., ingress and egress together). Either considering all the 74 transits or after excluding 4 light curves with $\sigma_\mathrm{T_0}$ $>$ 13 seconds, all our analyses of the $\beta$-PNR plane adopting $\sigma_\mathrm{T_0}$ as the perpendicular axis, did not show any clear correlation.

Symmetry can also play an important role in the determination of mid-transit times. Hence, we compared different quality indicators for symmetric sections of the light curve respect to $T_\mathrm{0}$: i) PNR between contact points 1 and 2 (ingress) and PNR between contact points 3 and 4 (egress), ii) $\beta$ between contact points 1 and 2 (ingress) and $\beta$ between contact points 3 and 4 (egress), iii) PNR/$\beta$ between contact points 1 and 2 (ingress) and PNR/$\beta$ between contact points 3 and 4 (egress), iv) $\chi^{2}$ of the model to data before $T_\mathrm{0}$ and after $T_\mathrm{0}$ (including, in both cases, the out-of-transit), and v) $\chi^{2}$ of the model to the ingress data-points and $\chi^{2}$ of the model to the egress data-points. No correlation between the degree of asymmetry, interpreted as the difference between quality indicators, and $\sigma_\mathrm{T_0}$ was found. We obtained the same null result in the search for a relation between the duration of the out-of-transit before/after ingress/egress.

In the absence of any correlation between the error in $T_\mathrm{0}$ and the quality or symmetry of the transit or the values of the LD coefficients, we decided to adopt $\sigma_\mathrm{T_0}$ i.e., the standard deviation found for $T_\mathrm{0}$ by varying the fixed values of the photometric parameters in $\pm 1 \sigma$, as the specific limit to the error in the mid-time of each transit due to the presence of systematics $e^{\beta}_\mathrm{T_0}$.

\section{TIMING ANALYSIS}\label{section:5}

The time stamps of the 74 full transits analyzed in this work were converted to Barycentric Julian Date in Barycentric Dynamical Time, $\mathrm{BJD}_\mathrm{TDB}$, with the online tool\footnote{http://astroutils.astronomy.ohio-state.edu/time/hjd2bjd.html} that uses the mathematical transformations described in \citet{eastman}.

The mid-transit time of each light curve was computed with the JKTEBOP code by performing 10\,000 Monte Carlo (MC) simulations and by using the residual permutation (RP) algorithm. In both cases, the only freely varying quantities were $\mathrm{T}_\mathrm{0}$, $l_{0}$, the coefficients of the polynomial to fit the out-of-transit data-points and the LD coefficients when corresponding. As initial values for the parameters, we adopted $i$, $k$, and $\Sigma$ as derived in Section \ref{section:3}, and for $e$, $P$, mass ratio, and the quadratic LD coefficients we kept the same quantities used in our first run of JKTEBOP. To be conservative, we adopted for the mid-time of each transit the mean value and the symmetric uncertainties, $\pm \sigma$, given by the task (MC or RP) that provided the largest error. Here, $\pm \sigma$ or equivalently $e^{f}_{T0}$, correspond to the 68.3$\%$ values (or the 15.87th and 84.13th percentiles) of the selected distribution. In order to provide reliable uncertainties, the final error for each mid-transit time measurement was computed as:

\begin{equation}
     (e_\mathrm{T0})^2 = (e^{f}_\mathrm{T0})^2 + (e^{\beta}_\mathrm{T0})^2,
\end{equation}

\noindent where $e^{f}_{T0}$ is the formal error calculated by the fit and $e^{\beta}_\mathrm{T_0}$ is the systematic error that accounts for the influence of the adopted values of the photometric parameters on the derived mid-times (see previous section). Our measurements of the mid-transit times of each of the 74 full transits are given in Table \ref{table:6}.

\begin{table}
\renewcommand\thetable{6}
\caption{Mid-transit times and uncertainties calculated in this work}             
\label{table:6}      
\centering        
\begin{tabular}{ccccc}     
\hline\hline       
Epoch	&	$\mathrm{T}_{0}-2450000$	& $e_\mathrm{T0}$ &	$e^{f}_\mathrm{T0}$	& $e^{\beta}_\mathrm{T0}$\\
	&	[$\mathrm{BJD}_\mathrm{TDB}$]	& [s] &	[s]	& [s]\\
\hline         
53	&	4817.146361	&	13.95	&	13.88	&	1.30	\\
609	&	5255.741162	&	8.68	&	8.63	&	0.96	\\
614	&	5259.684594	&	31.02	&	31.02	&	0.19	\\
632	&	5273.882529	&	61.78	&	61.35	&	7.28	\\
709	&	5334.626037	&	12.98	&	12.89	&	1.52	\\
714	&	5338.569384	&	21.14	&	19.44	&	8.29	\\
733	&	5353.557749	&	13.21	&	13.09	&	1.77	\\
752	&	5368.545551	&	39.68	&	39.00	&	7.33	\\
969	&	5539.722944	&	18.00	&	17.95	&	1.31	\\
1007	&	5569.698234	&	16.33	&	16.32	&	0.77	\\
1021	&	5580.741547	&	22.09	&	22.01	&	1.87	\\
1026	&	5584.686886	&	16.14	&	16.03	&	1.90	\\
1026	&	5584.686843	&	13.94	&	13.92	&	0.71	\\
1038	&	5594.152340	&	50.18	&	50.04	&	3.77	\\
1047	&	5601.252355	&	35.19	&	35.12	&	2.27	\\
1049	&	5602.831343	&	39.20	&	38.66	&	6.46	\\
1054	&	5606.774231	&	15.51	&	15.50	&	0.67	\\
1055	&	5607.562499	&	25.35	&	25.30	&	1.63	\\
1074	&	5622.550599	&	23.09	&	22.26	&	6.11	\\
1076	&	5624.128431	&	46.92	&	46.37	&	7.16	\\
1116	&	5655.682291	&	36.15	&	36.13	&	1.11	\\
1159	&	5689.603433	&	36.93	&	36.27	&	6.96	\\
1164	&	5693.547077	&	13.11	&	13.09	&	0.83	\\
1178	&	5704.590961	&	26.20	&	26.18	&	1.07	\\
1183	&	5708.535638	&	13.56	&	13.56	&	0.35	\\
1552	&	5999.616340	&	24.83	&	24.82	&	0.84	\\
1552	&	5999.616494	&	17.38	&	17.33	&	1.26	\\
1552	&	5999.616359	&	10.87	&	10.84	&	0.80	\\
1552	&	5999.616055	&	16.78	&	16.64	&	2.15	\\
1552	&	5999.616139	&	25.82	&	25.78	&	1.45	\\
1552	&	5999.616338	&	17.71	&	17.68	&	1.10	\\
1552	&	5999.616365	&	11.07	&	11.07	&	0.12	\\
1552	&	5999.616509	&	18.98	&	18.98	&	0.27	\\
1552	&	5999.616751	&	28.73	&	28.67	&	1.82	\\
1580	&	6021.703582	&	13.40	&	13.10	&	2.82	\\
1580	&	6021.703343	&	14.33	&	14.32	&	0.60	\\
1580	&	6021.703628	&	9.85	&	9.79	&	1.09	\\
1580	&	6021.704079	&	10.46	&	10.37	&	1.37	\\
1580	&	6021.703801	&	11.05	&	11.02	&	0.77	\\
1580	&	6021.703225	&	15.06	&	15.04	&	0.88	\\
1580	&	6021.703927	&	10.32	&	10.31	&	0.49	\\
1580	&	6021.703713	&	12.02	&	11.98	&	0.98	\\
1580	&	6021.704289	&	41.68	&	41.59	&	2.65	\\
1595	&	6033.538366	&	24.69	&	24.67	&	0.94	\\
1595	&	6033.537996	&	34.28	&	33.62	&	6.65	\\
1595	&	6033.537131	&	63.61	&	60.67	&	19.12	\\
1633	&	6063.511736	&	21.39	&	19.75	&	8.20	\\
1964	&	6324.620530	&	52.31	&	52.31	&	0.24	\\
1977	&	6334.872072	&	44.81	&	44.72	&	2.77	\\
2048	&	6390.880326	&	46.21	&	43.88	&	14.48	\\
2106	&	6436.632877	&	56.16	&	56.16	&	0.27	\\
2465	&	6719.826225	&	47.92	&	47.72	&	4.27	\\
2490	&	6739.547650	&	13.19	&	13.06	&	1.83	\\
2490	&	6739.547358	&	26.13	&	26.13	&	0.06	\\
2792	&	6977.776470	&	9.07	&	6.30	&	6.53	\\
3351	&	7418.736819	&	20.20	&	20.20	&	0.17	\\
3389	&	7448.712917	&	6.62	&	6.59	&	0.62	\\
3477	&	7518.131796	&	32.48	&	32.48	&	0.47	\\
3513	&	7546.529748	&	23.84	&	23.84	&	0.47	\\
3518	&	7550.472701	&	38.68	&	37.46	&	9.65	\\
3551	&	7576.504698	&	64.34	&	64.08	&	5.77	\\
\hline                  
\end{tabular}
\end{table}

\begin{table}
\renewcommand\thetable{6}
\caption{Continued}             
\label{table:6}      
\centering        
\begin{tabular}{ccccc}     
\hline\hline       
Epoch	&	$\mathrm{T}_{0}-2450000$	& $e_\mathrm{T0}$ &	$e^{f}_\mathrm{T0}$	& $e^{\beta}_\mathrm{T0}$\\
	&	[$\mathrm{BJD}_\mathrm{TDB}$]	& [s] &	[s]	& [s]\\
\hline         

3739	&	7724.807825	&	71.55	&	71.23	&	6.79	\\
3758	&	7739.794151	&	46.36	&	45.12	&	10.64	\\
3806	&	7777.657426	&	121.27	&	120.50	&	13.67	\\
3815	&	7784.758917	&	15.19	&	15.07	&	1.89	\\
3825	&	7792.647526	&	65.59	&	65.54	&	2.64	\\
3830	&	7796.591396	&	27.26	&	27.22	&	1.38	\\
3906	&	7856.542283	&	45.30	&	45.29	&	1.05	\\
4189	&	8079.785673	&	18.51	&	18.33	&	2.51	\\
4227	&	8109.763837	&	131.92	&	131.83	&	4.64	\\
4359	&	8213.887473	&	21.01	&	20.89	&	2.23	\\
4757	&	8527.845759	&	39.13	&	38.98	&	3.52	\\
4784	&	8549.144717	&	46.65	&	44.80	&	13.01	\\
4829	&	8584.641667	&	23.14	&	23.13	&	0.37	\\
\hline                  
\end{tabular}
\end{table}

\subsection{Searching for possible orbital decay}

In order to confirm or rule out a possible orbital decay in the system, we fitted a linear and a quadratic model to the mid-transit times as a function of epoch. The linear model assumes that the orbit is circular and the orbital period, $P$, is constant:

\begin{equation}
\mathrm{T}_{\mathrm{0}}(E)=\mathrm{T}_\mathrm{ref} +  E \times P
 \label{efe:lin}
\end{equation}

\noindent where $E$ is the epoch number and $\mathrm{T}_\mathrm{ref}$ a reference minimum time. The quadratic model also assumes a circular orbit but the orbital period changes at a steady rate, $dP/dE$:

\begin{equation}
\mathrm{T}_{\mathrm{0}}(E)=\mathrm{T}_\mathrm{ref} +  E \times P + \frac{1}{2} \times \frac{dP}{dE} \times E^{2}.  
 \label{efe:cua}
\end{equation}

\begin{figure*}
   \centering
   \includegraphics[width=.6\textwidth]{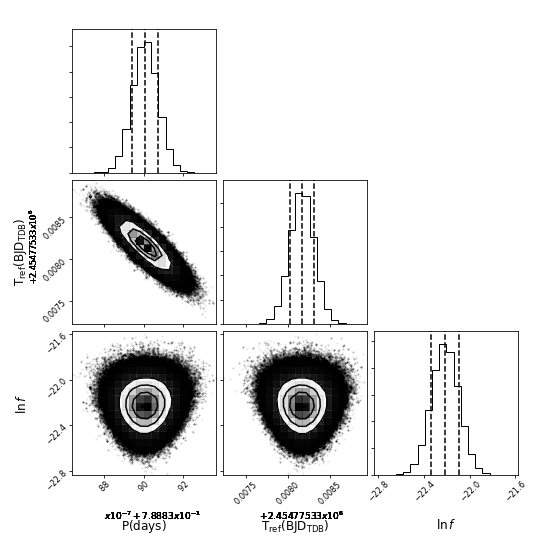}
   \includegraphics[width=.65\textwidth]{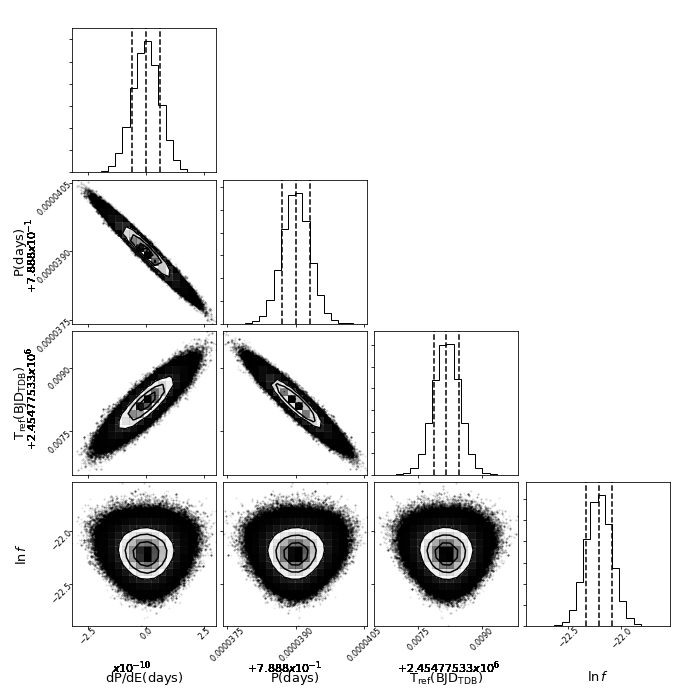}
   \caption{MCMC posterior probability distributions of the free parameters adopted for the linear (top panel) and quadratic (bottom panel) models. Dashed lines show the median and 16$\%$ and 84$\%$ percentiles of each distribution. In both cases, the variable \textit{f} is a fractional amount that takes into account any possible underestimation in the variance of the data.}   
      \label{corner}
\end{figure*}

\noindent We fitted each model by assuming a Gaussian likelihood function and used the \textit{emcee} MCMC sampler implementation of \citet{foreman} to sample over the posterior probability distributions for all the free parameters (Figure \ref{corner}), i.e. $\mathrm{T}_\mathrm{ref}$ and $P$ in Eq. \ref{efe:lin} and $\mathrm{T}_\mathrm{ref}$, $P$, and $dP/dE$ in Eq. \ref{efe:cua}. The prior corresponding to the changing orbital period in the quadratic ephemeris was allowed to vary between positive and negative numbers.   

\begin{table*}
\renewcommand\thetable{7}
\caption{Best-fit parameters for the linear and quadratic models}             
\label{table:7}      
\centering          
\begin{tabular}{cccc}     
\hline\hline       
 Model & $T_\mathrm{ref}$[$BJD_\mathrm{TDB}$] & $P$[days] & $dP/dE$[days]\\
\hline                    
Linear & $2454775.33817^{+0.00014}_{-0.00014}$ & $0.788839007^{+0.000000064}_\mathrm{-0.000000064}$ &  -- \\
Quadratic & $2454775.33817^{+0.00030}_{-0.00030}$  & $0.78883900^{+0.00000030}_{-0.00000029}$	  & $(0.89972 \times 10^{-12})^{+0.000000000058}_{-0.000000000058}$  \\
\hline                  
\end{tabular}
\end{table*}

In Table \ref{table:7}, we present the values and uncertainties adopted from the median and 16$\%$ and 84$\%$ percentiles of the drawn posterior distributions for the fitted parameters in the linear and quadratic models. Through the relation $dP/dt=P^{-1}(dP/dE)$, it is straightforward to calculate the change in the period over time, $dP/dt$. In this case,

\begin{equation}
\dot{P}=1.14(+73)(-74) \times 10^{-12}=0.036^{+2.3}_{-2.33} \quad \textrm{ms yr}^{-1}
\end{equation}

 \noindent is essentially consistent with zero. The comparison between the reduced chi-square values and the residuals root mean square of the best-fits for the linear  ($\chi^{2}_\mathrm{r}=6.31$, 72 degrees of freedom and $rms=$1.19 minutes) and quadratic ($\chi^{2}_\mathrm{r}=6.35$; 71 degrees of freedom and $rms=$1.19 minutes) models indicates that the linear ephemeris provides a marginally better representation of the data than the quadratic one. Two other useful and widely used metrics to decide which model better represents the data are the Bayesian Information Criterion (BIC) and the Akaike Information Criterion (AIC), defined as $\mathrm{BIC} = \chi^{2} + k_\mathrm{F}\log{N_\mathrm{P}}$ and $\mathrm{AIC} = \chi^{2} + 2k_\mathrm{F}$, respectively, where $k_\mathrm{F}$ is the number of free parameters in the model and $N_\mathrm{P}$ is the number of data-points (74 in this analysis). For the linear fit ($k_\mathrm{F}=$2), we measured $BIC_\mathrm{lin}=$ 463.04 and $AIC_\mathrm{lin}=$ 458.43, and $BIC_\mathrm{quad}=$ 464.05 and $AIC_\mathrm{quad}=$ 457.14 for the quadratic case ($k_\mathrm{F}=$3). The difference in the BIC values between both models is $\Delta BIC=BIC_\mathrm{quad}-BIC_\mathrm{lin}=$1.01, which according to \citet{kass} favors the constant period model over the changing orbital period scenario. On the contrary, for the AIC values, $\Delta AIC=AIC_\mathrm{lin}-AIC_\mathrm{quad}=$1.29 gives support to the quadratic model over the linear one \citep{burnham}. Here, it is important to notice that for both, $\Delta BIC$ and $\Delta AIC$, evidence in favor of one or another model is not strong. The reason for this discrepancy mainly arises because the quadratic coefficient is so small that both models can be considered as representative of a linear ephemeris. 
 
 Moreover, a global analysis of all these results, summarized in Table \ref{table:8} and shown in Figure \ref{oc}, also seems to indicate that the evolution of the mid-transit times of WASP-19b over 10 years is best explained by a constant orbital period. This conclusion contradicts predictions of detectable variation in the mid-transit times after a decade of observations \citep{valsechi, essick}. The seemingly best quadratic scenario predicts a very slow change in period, which would require $\sim 3 \times 10^{5}$ years of observations to detect a shift in mid-transit times of 10 seconds. Furthermore, this solution has a positive quadratic coefficient, indicating an increase in orbital period and not a decrease as expected due to tidal decay. Although the quadratic term is greater than zero and the evidence supporting a change in period is not strong, it is still possible to set an upper limit on the orbital decay of $\dot{P}_\mathrm{up}=-2.294$ ms $yr^{-1}$, by computing $\dot{P}_\mathrm{up}=\dot{P}-\sigma^{-}_{\dot{P}}$, where $\sigma^{-}_{\dot{P}}$ is the negative error found for the steady change in $P$. 

The dimensionelss quantity $Q'_{\star}$, called modified tidal quality factor and defined in terms of the maximum tidal energy stored in the system relative to the energy dissipated in one cycle \citep{goldreich}, is very important to characterize the evolution of the system under the effect of tidal forces. Specifically, large values of $Q'_{\star}$ imply inefficient tidal dissipation and hence a slow orbital evolution, while small values of $Q'_{\star}$ correspond to an efficient tidal dissipation resulting in a fast orbital evolution. It can be estimated by taking the upper limit of $\dot{P}$ through the expression:

\begin{equation}
   Q'_{\star}= \frac{-27 \pi}{2 \dot{P}} \Big(\frac{M_\mathrm{P}}{M_\mathrm{\star}}\Big)  r_\mathrm{\star}^{5},
   \label{coeficiente_tidal}
\end{equation}

\noindent obtained by rearranging the Eq. 5 of \citet{wilkins} in terms of $\dot{P}$. In this study, the values of the planetary and stellar masses, $M_\mathrm{P}$ and $M_\mathrm{\star}$, were adopted from \citet{hellier}, and the value assumed for the stellar radius relative to the semimajor axis, $r_\mathrm{\star}$, was the one determined in Section \ref{section:3}. The uncertainties in $Q'_{\star}$ were computed by propagating the errors of $M_\mathrm{P}$, $M_\mathrm{\star}$, and $r_\mathrm{\star}$. Thus, we derived a lower limit on the modified quality factor of:

\begin{equation*}
    Q'_{\star} > (1.23 \pm 0.231) \times 10^{6}.     
\end{equation*}

\begin{figure}
   \centering
   \includegraphics[width=.5\textwidth]{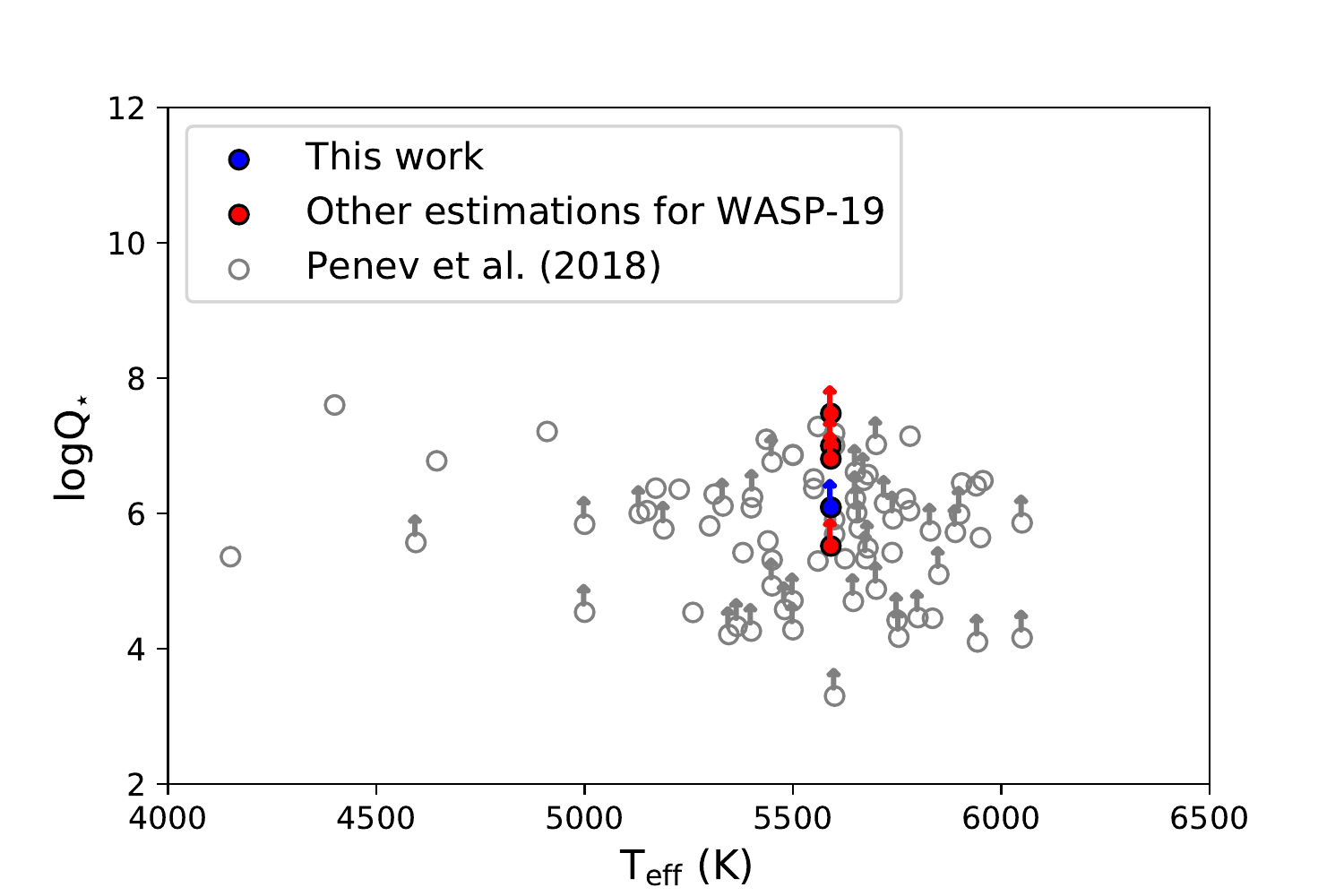}
   \caption{$T_\mathrm{eff}$ versus $\log Q'_{\star}$. The blue circle marks the value of $\log Q'_{\star}$ computed in this work for WASP-19, while red symbols indicate the results for the same target obtained in previous studies \citep{hebb, brown, abe, penev}. As comparison, the values calculated by \citet{penev} for 75 other planets are presented in gray empty circles. Arrows pointing up mark lower limits of $Q'_{\star}$. For a better visualization, error bars in $T_\mathrm{eff}$ are not shown.}
   \label{teffvslogq}
\end{figure}

\noindent In a general context, our estimation of $Q'_{\star}$ is in the wide range of expected values (between $10^{5}-10^{9}$) predicted by studies based on large samples of hot Jupiters \citep{jackson, essick, bonomo, penev, collierjar, hamer}. If compared with specific determinations of the modified tidal quality factor for WASP-19, our measurement is also in agreement with previous estimates \citep{hebb, brown, abe, penev}. These comparisons can be visualized in Figure \ref{teffvslogq}, where we plot $T_\mathrm{eff}$ versus $log(Q'_{\star})$ for WASP-19 (in which the blue circle is the result obtained in this work and the red ones are the measurements of previous studies) and 75 other planets indicated in gray empty circles, for which \citet{penev} derived $Q'_{\star}$ by modeling the orbital and the stellar spin evolution.

\begin{table*}
\renewcommand\thetable{8}
\caption{Comparison between the linear and quadratic model}             
\label{table:8}      
\centering          
\begin{tabular}{ccccc}     
\hline\hline       
 Model & $\chi^{2}_\mathrm{r}$ & $rms$[minutes] & $BIC$ & $AIC$\\
\hline                    
Linear & 6.31 & 1.19 & 463.04 & 458.43 \\
Quadratic & 6.35 & 1.19 & 464.05 &  457.14\\
\hline                  
\end{tabular}
\end{table*}

\subsection{Searching for possible companions}\label{oc}

\citet{mancini} and \citet{espinoza} found that a linear ephemeris is not a good fit to the data of WASP-19b. Then, to assess if these findings can be the result of perturbations produced by another body gravitationally bound to the system, we investigated a possible sinusoidal variation in the mid-transit times. To check this out, we searched for periodicities in the O-C residuals (see Figure \ref{oc}), computed as the difference between the $\mathrm{T}_\mathrm{0}$ values estimated with JKTEBOP and those predicted by the linear ephemeris calculated in the previous section, by running a Lomb-Scargle (LS) periodogram \citep{horne} and a Phase Dispersion Minimization (PDM) algorithm \citep{pdm} to the data. The LS routine encountered a peak at 31.35 epochs or, equivalently, a period $\sim$ 25 days with an $FAP=53 \%$ estimated through 1000 Monte Carlo simulations, while PDM found a period around 198 days with $\Theta$ $\sim$ 0.86. The discrepancy in the periods found with both tasks and the high values of $FAP$ and $\Theta$ are indicative of a null detection of periodic variations in the mid-transit times. However, as can be seen in the next section, it is still possible to set an upper limit on the mass of a potential perturber (i.e. another planetary-mass body) bounded to the system, capable of producing detectable periodic transit timing variations (TTVs) in our data, by using the standard deviation found in the O-C residuals ($\sigma_\mathrm{TTV}=1.19$ minutes or, equivalently, 71.4 seconds).   

\begin{figure*}
   \centering
   \includegraphics[width=.8\textwidth]{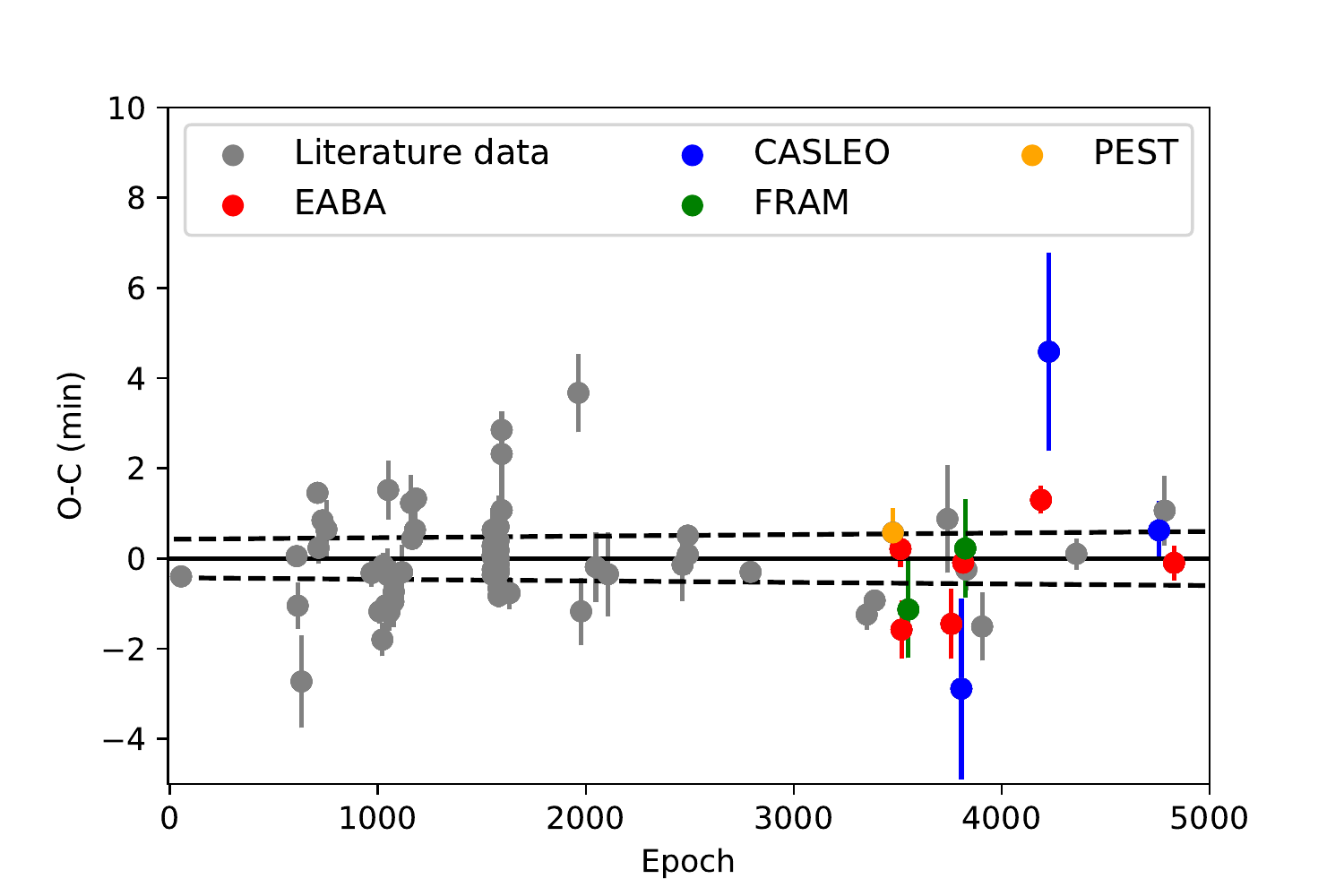}
      \caption{O-C data-points versus Epoch considering the 74 complete transit light curves analyzed in this study. Different colors indicate transits extracted from the literature (gray circles) and red, blue, green, and orange circles point out the transits observed by our team at different facilities (see the legend). Dashed black lines represent $\pm 2\sigma$ errors in the linear ephemeris. Some timing measurements have significant deviations from the linear ephemeris that might be caused by mutual close-encounters between the transiting and a non detected planet. However, this possibility certainly deserves further investigation which is beyond the scope of this paper.}
         \label{oc}
   \end{figure*}

\subsection{Constraints on the mass of a possible perturber}

Measuring transit-timing variations provides a method to detect additional massive bodies that might be
present (possibly non-transiting) in the system. In principle, sufficiently strong gravitational
perturbations by the unseen companion, will cause the mid-transit time of the transiting planet to change periodically over time
\citep{agol2005,holman2005,nesvorny2008} with a given amplitude depending on the perturbing bodies mass and orbital parameters. The TTV amplitude can be quantified by the root-mean-square (RMS) statistic which measures the scatter of mid-transit timing data around the nominal (unperturbed) linear ephemeris. Qualitatively, larger perturbations would result in a larger RMS scatter around the nominal ephemeris. For orbital architectures involving mean-motion resonances the TTV effect is amplified \citep{agol2005,holman2005} and allows, in principle, the detection of very low-mass perturbing objects. For sufficiently precise timing data the modelling of the TTV signal enables one to infer the mass and 
orbit of the perturbing body \citep{Holman2010,Ballard2011}. In this work we will not carry out a detailed TTV modeling analysis and will use the empiricially measured RMS scatter to infer an estimate of the upper mass limit of the perturbing body.

Here we apply the technique for TTV calculations as described in \citet{TEMP2,TEMP4,TEMP1}. The calculation of
an upper mass-limit is performed numerically from directly integrating the orbits. We have modified the
\texttt{FORTRAN}-based \texttt{MICROFARM}\footnote{\url{https://bitbucket.org/chdianthus/microfarm/src}}
package \citep{go2003,go2008} which utilizes OpenMPI\footnote{\url{https://www.open-mpi.org}} to spawn
hundreds of single-task parallel jobs on a suitable super-computing facility. The package main purpose is the
numerical computation of the Mean Exponential Growth factor of Nearby Orbits  
\citep[MEGNO]{cincsimo2000,go2001,ci2003} over a grid of initial values of orbital parameters for an $n$-body
problem. The calculation of the RMS scatter of TTVs in the present work follows a direct brute-force method,
which proved to be robust given the availability of computing power.

Within the framework of the three-body problem, we integrated the orbits of WASP-19b and an additional hypothetical perturbing planet. The mid-transit time was accurately calculated from an iteration process as a result of a series of back-and-forth integrations. The best-fit radii of
both the transiting planet and the host star were necessary input 
parameters. In general, initial conditions for the transiting planet (semi-major axis, eccentricity, argument of pericentre and mass), including stellar properties, were taken to be those presented in \citet{hellier}. The orbital inclination and the stellar and planetary radii were adopted from Table \ref{table:2}. The initial mean anomaly of the transiting planet was set to zero.

We then calculated an analytic least-squares regression to the time-series of transit epochs and mid-transit
times to determine a best-fitting linear ephemeris with an associated RMS statistic for the scatter of
computed transit times. The RMS statistic was based on a 20-year integration corresponding to $\simeq 9000$
transit events for WASP-19b. This procedure was then applied to a grid of masses and semi-major axes of the
perturbing planet. We chose to encode the semi-major axis as the orbital period using the third law of Kepler.

In principle, no information about the properties of the perturbing body is available, except for a possible TTV signal. In this study, we have chosen to start the perturbing planet for two different orbital eccentricities: $i)$ circular ($e=0$) and $ii)$ moderate eccentricity ($e=0.1$). The orbit of the perturbing planet is initially co-planar with the transiting planet. This implies that $\Omega_2=0^{\circ}$ and
$\omega_2=0^{\circ}$ for the circular case of the perturbing planet. This setting provides a most conservative estimate of the upper mass limit of a possible perturber \citep{bean2009,fukui2011,Hoyer2011,Hoyer2012}. We
refer the interested reader to \citet{TEMP4} where the authors have studied the effect on upper mass limit for various initial orbital parameters. For example, while the initial eccentricity would change the overall system stability and introduce higher-order mean-motion resonances, the effect on upper mass limit due to different initial phase of the peturbing planet is subtle. The results are shown in Fig.~\ref{megnottv}. For the WASP-19b system, the measured
transit-timing RMS scatter taking into account the 74 transit light curves analyzed in this work was $\rm TTV_{\rm RMS} \simeq 71\,\rm s$. Considering the circular case, an additional planet with a mass as low as $\simeq 2~M_{\oplus}$, $\simeq 2.5~M_{\oplus}$ and $\simeq 1.0~M_{\oplus}$ at the 1:3, 1:2 (interior) and 2:1 (exterior) mean-motion resonances could cause the observed RMS scatter. Hypothetical planets of $\simeq 3.5~M_{\oplus}$ and $\simeq 2~M_{\oplus}$ could cause the observed RMS scatter at the 5:3 and 3:1 exterior mean-motion resonances. 

For the perturbing planet started on a moderate $e=0.1$ eccentric orbit the results change slightly. For the 1:3 and 1:2 resonance the measured TTV RMS scatter could be caused by a planet of mass $\simeq 18~M_{\oplus}$ and
$\simeq 1~M_{\oplus}$, respectively. Due to the larger eccentricity the general instability area now engulfs the 5:3 exterior resonance. For the 2:1 resonance a mass of $\simeq 2~M_{\oplus}$ could cause the observed 
TTV scatter. Qualitatively, almost no change is observed for the 3:1 resonance allowing the same conclusion as for the circular case. Finally, the 4:1 resonance plays a role and a hypothetical planet of mass 
$\simeq 3~M_{\oplus}$ could explain the TTV RMS variations.

\subsection{Dynamical MEGNO maps and orbital resonances}

In order to calculate the location of mean-motion resonances, we have used the same code to calculate MEGNO 
on the same parameter grid. However, this time we integrated each initial grid point for 1000 years, allowing
this study to highlight the location of weak chaotic high-order mean-motion resonances. In short, MEGNO
quantitatively measures the degree of stochastic behaviour of a non-linear dynamical system and has been
proven useful in the detection of chaotic resonances \citep{go2001,hi2010}. In addition to the Newtonian
equations of motion, the associated variational equations of motion are solved simultaneously allowing the
calculation of MEGNO at each integration time step. The \texttt{MICROFARM} package implements the
\texttt{ODEX}\footnote{\url{https://www.unige.ch/~hairer/prog/nonstiff/odex.f}} extrapolation algorithm to
numerically solve the system of first-order differential equations.

\begin{figure}
\includegraphics[width=1.0\columnwidth]{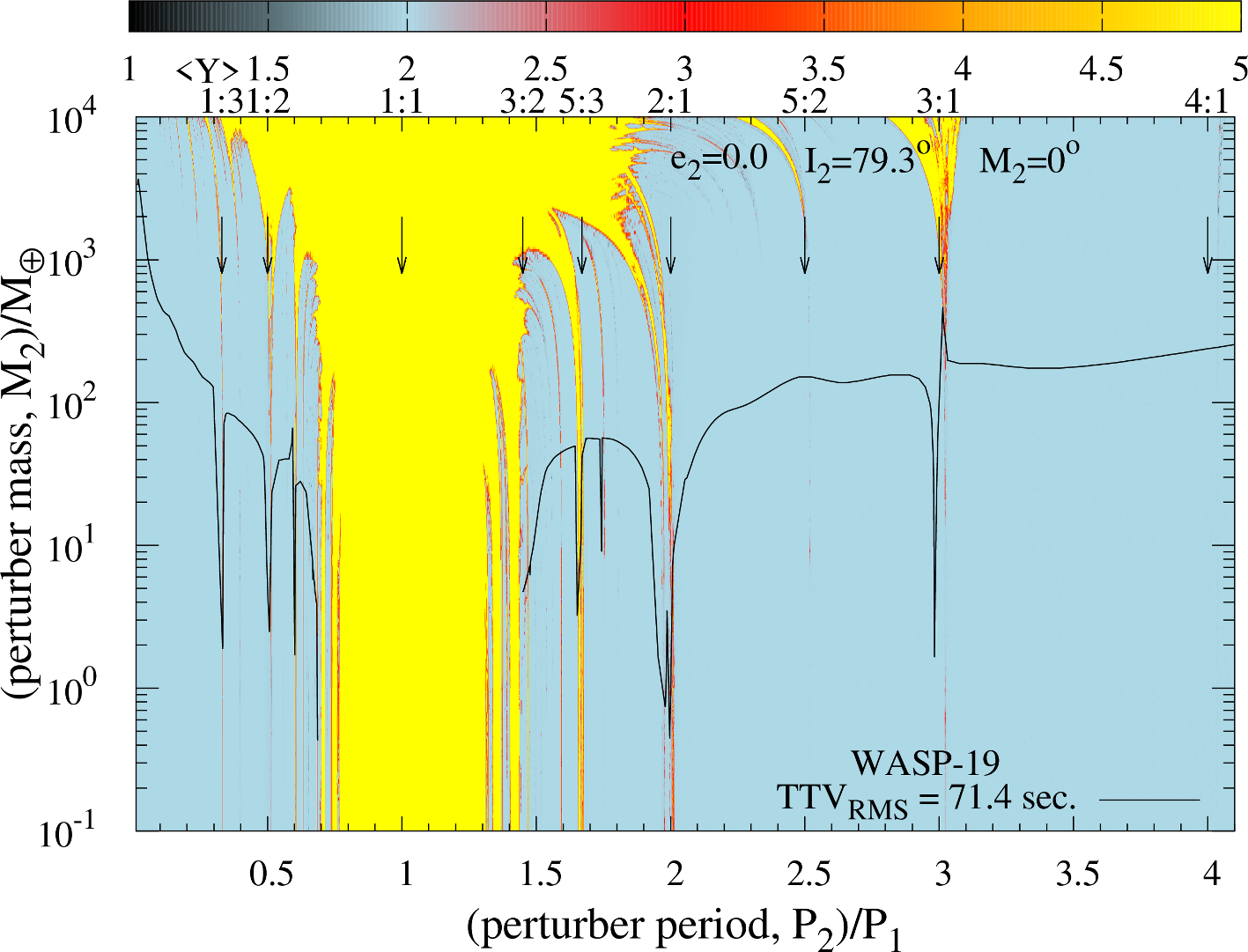}
\includegraphics[width=1.0\columnwidth]{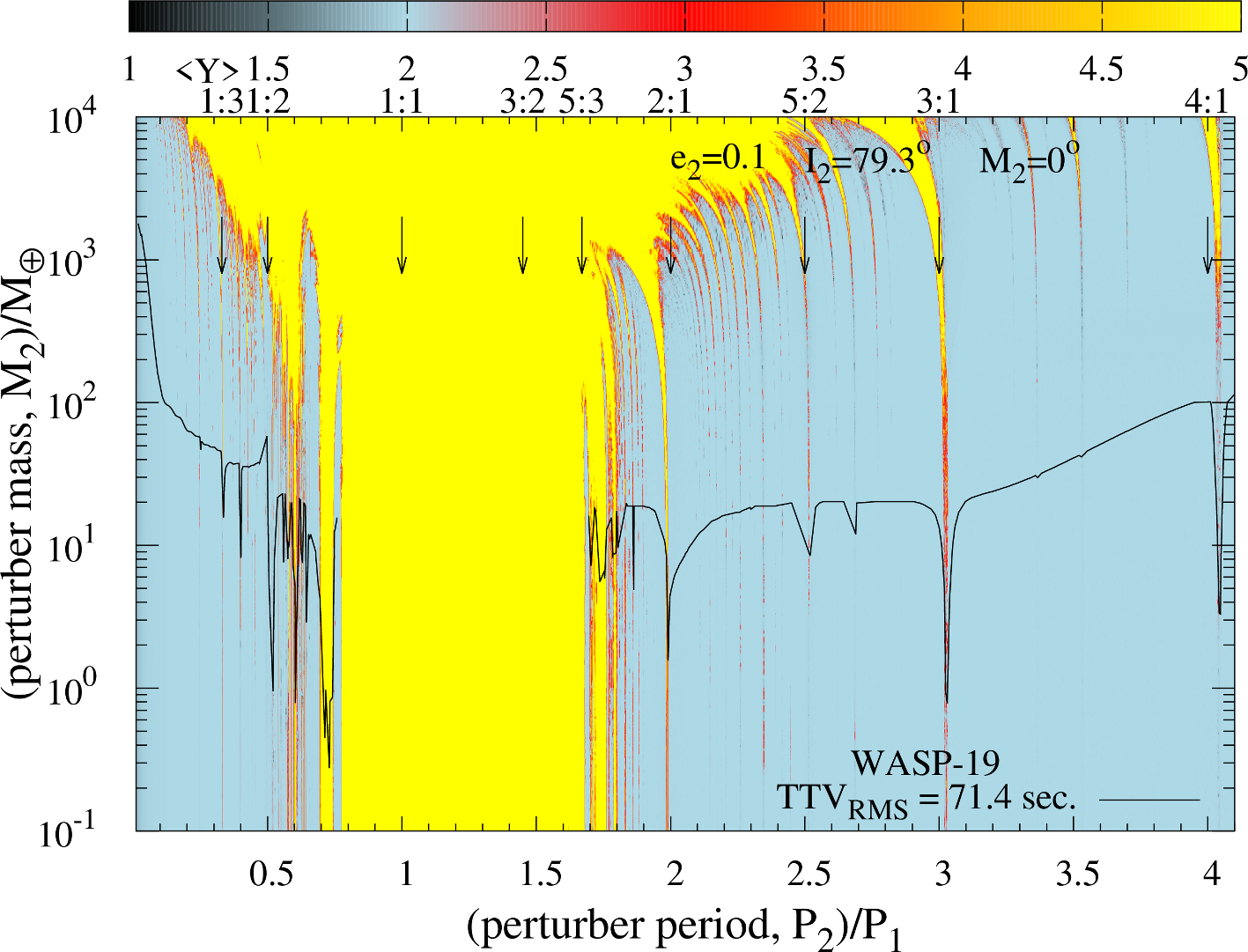}
\caption{Dynamical stability maps ($\langle Y \rangle$, MEGNO) for the WASP-19 transiting system considering the three-body problem. The parameters for the transiting planet were held fixed to the best-fit values. Quasi-periodic (i.e., bounded time evolution) initial conditions ($\langle Y\rangle$ close to 2.0) we chose to color-code MEGNO with a blue color. For initial conditions leading to chaotic (i.e., irregular time evolution) dynamics the MEGNO index is diverging away from 2.0 and we chose to color code with yellow. The black solid line shows the upper mass-limit of a perturbing body which produces a root-mean-square ($\rm TTV_{\rm RMS}$) transit-timing variation of $\simeq 71$ s as a function of period (or semi-major axis). The vertical arrows indicate orbital resonances between the two bodies. We considered two different initial eccentricities $e = 0.0$ (top panel) and $e = 0.1$ (bottom panel) of the perturbing planet. The two planets were assumed to be on initial co-planar orbits. \emph{See electronic version for colors}.}
\label{megnottv}
\end{figure}

Following the definition of MEGNO \citep{cincsimo2000}, denoted as $\langle Y\rangle$ in
Fig.~\ref{megnottv}, in a dynamical system that evolves quasi-periodically, the
quantity $\langle Y \rangle$ will asymptotically approach 2.0 for $t \rightarrow \infty$. In that case, often
the orbital elements associated with that orbit are bounded to within a certain range with no sign of diffusion in their time evolution. In case of a chaotic time evolution, the $\langle
Y\rangle$ diverges away from 2.0 with orbital parameters exhibiting erratic temporal excursions. For
quasi-periodic orbits, we typically have $|\langle Y \rangle - 2.0| < 0.001$ at the end of each integration. 

Importantly, MEGNO is unable to prove that a dynamical system is evolving quasi-periodically, meaning that a
given system cannot be proven to be stable or bounded for all times. The integration of the equations of
motion only considers a limited time period. However, once a given initial condition has found to be chaotic,
there is no doubt about its erratic nature in the future.

From Fig.~\ref{megnottv} we find the usual instability region located in the proximity of the transiting planet ($P_2/P_1 \simeq 1.0$; where $P_2$ and $P_1$ are the orbital periods of the perturber and WASP-19b, respectively) with MEGNO color-coded as yellow (corresponding to $\langle Y\rangle > 5$). 
The extent of this region coincides with the results presented in \citet{barnes2006}. In each map the
locations of mean-motion resonances are indicated by vertical arrows.

\section{SUMMARY AND DISCUSSION}\label{section:6}

We have performed the first empirical study of orbital decay in the exoplanet WASP-19b, through the homogeneous determination of the mid-transit times of 74 complete transit light curves covering over 10 years. Contrary to what is expected by theoretical predictions (e.g., \citealt{valsechi, essick}), we found no evidence of orbital decay nor periodic variations in the transit timings that would imply the presence of other bodies in the system. Nonetheless, we were able to put constraints on the upper mass of a possible perturber by carrying out a dynamical analysis. Particularly, at the first order mean-motion resonances, 1:2 and 2:1, we can exclude planetary companions of a few $M_{\earth}$ in mass, in both circular and moderate eccentric orbits.

A linear ephemeris is the best representation of the evolution of the mid-transit times as a function of epoch. Our analysis allowed us to place an upper limit on the steady changing rate of the orbital period, $\dot{P}=-2.294$ ms $yr^{-1}$, and thus a lower limit for the modified tidal quality factor, $Q'_{\star} = (1.23 \pm 0.231) \times 10^{6}$. Both quantities in good agreement with previous estimations (Figure \ref{teffvslogq}). So far, the more promising candidates for presenting TTVs detectable with ground-based telescopes are the planetary systems WASP-4 \citep{wilson} and WASP-12 \citep{hebbw12}. For the first one, \citet{bouma} found that the transits observed by the TESS mission \citep{ricker} occurred 82 seconds earlier than  predicted by the linear ephemeris. Later, \citet{souw4} confirmed this variation through the analysis of 22 new transits of WASP-4b. Both tidal orbital decay and apsidal precession were proposed as plausible explanations for the transit timing variations detected, but these results are not conclusive yet. For WASP-12, \citet{maciew12, macie} measured a departure from a constant period consistent with orbital decay. This shortening in the orbital period has been detected by \citet{patra} with observations of new transits and occultations, although these authors also proposed other two scenarios, different from orbital decay, to explain their measurements: the existence of a long-period third body in the system and orbital precession. Recently, \citet{baluev} performed a homogeneous analysis of the transits data of both systems and obtained contradicting results. On one hand, for WASP-4b, they do not find any TTV signal and indicate that the finding reported by previous studies appears to be model-dependent. On the other hand, for WASP-12b, they confirm with high significance the nonlinear trend detected in previous works but conclude that 10$\%$ of the observed TTV is produced by light-travel effect which dilutes the strength of the rate of orbital decay previously measured. Given the lack of clear empirical signs of orbital decay and the difficulty to theoretically estimate the stellar modified tidal quality factor, it is very important to perform long-term monitoring of those systems with short-period giant planets. In this context, even a negative result such as that found in this study becomes relevant to shed light on how planetary systems evolve due to stellar tides. If WASP-19 is observed by the TESS and CHEOPS \citep{broeg} missions, future transits of this exoplanet, and hence new precise measurements of the mid-times will be acquired. These new estimations, added to those presented in this paper in Table \ref{table:6}, will enable a revision to the values of $\dot{P}$ and $Q'_{\star}$ determined in this study.

One of the most accepted theories on tides is that based on the work by \citet{zahn75, zahn77} that explored the contribution of different physical mechanisms on the tidal friction of close binary stars. This work found that for stars with convective cores and radiative envelopes (i.e., those with masses and effective temperatures larger than approximately those of the Sun), radiative damping is the dominant mechanism, leading to a less effective dissipation of tidal energy than in stars with radiative cores and convective envelopes, where turbulent dissipation is dominant. Therefore, it might be expected to measure large values of $Q'_{\star}$ in more massive and hotter than the Sun host stars and small values in hosts less massive and cooler than the Sun. Several subsequent studies \citep{barker, essick, penev}, explored different regimes of this theory applied to exoplanetary systems where the hosts are main-sequence or slightly evolved stars, resulting in a range of seven orders of magnitude for the possible values of the stellar modified tidal quality factor ($10^{5}-10^{12}$). Thus, in order to move forward in our understanding of this matter, it is necessary to obtain empirical model-independent estimations of $Q'_{\star}$ derived directly from observational data, as done in this paper.

\begin{figure*}
   \centering
   \includegraphics[width=.8\textwidth]{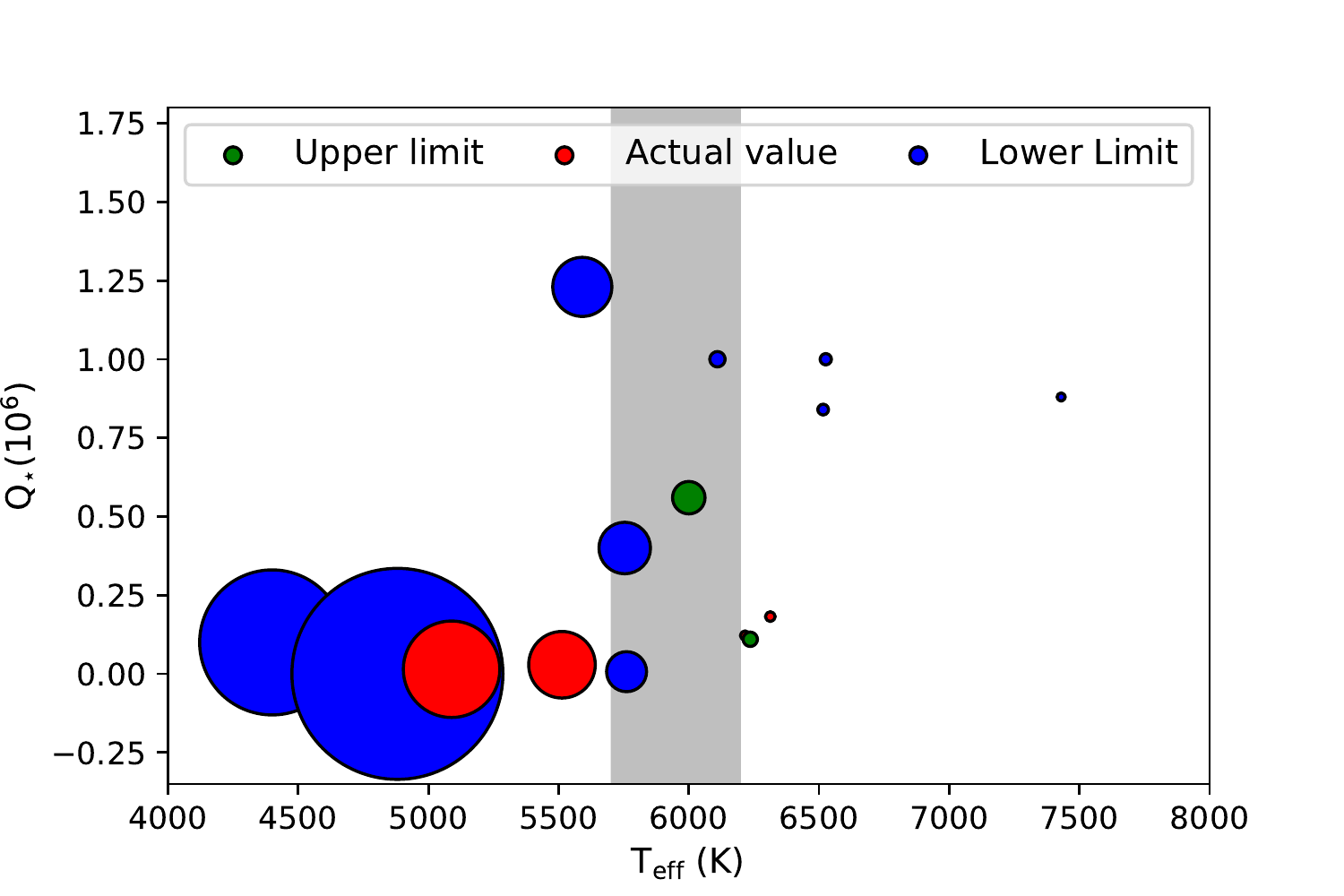}
      \caption{$T_\mathrm{eff}$ versus $Q'_{\star}$ for the 15 systems presented in Table \ref{table:9}. Red, blue, and green colors indicate measured actual values, and measured lower and upper limits of $Q'_{\star}$, respectively. The size of the circles scales with the percentage of the stellar convective zone mass respect to the total mass of the star, where the biggest circle corresponds to K2-39 with $m_\mathrm{CZ}=46\%$ and the smallest one to WASP-33 with $m_\mathrm{CZ}=0.000004\%$. The gray shaded area marks the range of effective temperatures at which $m_\mathrm{CZ}$ changes from 2$\%$ to 0.035$\%$, according to the models of \citet{baraffe}. For a better visualization, error bars in $T_\mathrm{eff}$ are not shown.} 
   \label{teffvsq}
   \end{figure*}

In this context, we tested the predictions by Zahn with a sample of 15 systems composed by a star and a planet, with measured values of the modified tidal quality factor. In all the cases, these $Q'_{\star}$ values were empirically determined by fitting the mid-transit times, measured from observed transits, as a function of epoch with a quadratic ephemeris. In Figure \ref{teffvsq}, we show a plot of effective temperature versus stellar modified tidal quality factor for these 15 exoplanetary systems. Here, green circles represent values of $Q'_{\star}$ that have been measured to be upper limits; blue circles are measured lower limits of $Q'_{\star}$; and red circles are measured $Q'_{\star}$ values. The gray shaded area can be interpreted as a transition zone that distinguishes stars with significant convective envelopes (below the left limit) from stars with negligible convective envelopes (beyond the right limit).  
Specifically, this area indicates the range of effective temperatures at which the percentage of the stellar convective zone mass ($M_\mathrm{CZ}$) respect to the total mass of the star i.e., $m_\mathrm{CZ}=(M_\mathrm{CZ}/M_{\star}) \times 100$, changes from 2$\%$ ($M_{\star} \sim$ 1 $M_\mathrm{\sun}$ and $T_\mathrm{eff} \sim$ 5700 K) to 0.035$\%$ ($M_{\star} \sim$ 1.4 $M_\mathrm{\sun}$ and $T_\mathrm{eff} \sim$ 6200 K), according to the models of \citet{baraffe} considering an age of $\sim$ 3 Gyr.

As can be seen, this plot suggests that stars with effective temperatures above $\sim$ 5600 K present $Q'_{\star}$ measurements that are larger than $\sim 1.12 \times 10^{5}$, meanwhile stars with effective temperatures down this limit have $Q'_{\star}$ values below this threshold. Our result implies that the change in the structure of the host star (from convective envelope/radiative core to radiative envelope/convective core), that would occur $\sim$ 6200 K considering the upper limit of the shaded region, is not the crucial factor affecting how quick the tidal evolution of its companion could happen. What might be important is the percentage of the stellar mass in the convective envelope, $m_\mathrm{CZ}$, given in Figure \ref{teffvsq} by the size of the points. To be consistent, we used Eq. 5 of \citet{murray} to homogeneously estimate $M_\mathrm{CZ}$ for the main-sequence and slightly evolved stars analyzed, while for the only giant star in our sample, K2-39 \citep{vaneylen}, we assumed a convective mass of 0.7 $M_\mathrm{\sun}$ \citep{donascimento, pasquini}. In this figure, it is quite clear that the stars with more massive convective zones, and also $T_\mathrm{eff}$ < 5600 K, tend to present values of $Q'_{\star} < 1.12 \times 10^{5}$ and viceversa for stars with smaller $m_\mathrm{CZ}$ (below $\sim 4 \%$). Then, fast orbital changes (small $Q'_{\star}$) due to tidal forces are expected for the planets around stars with significant convective zones, and slow evolution (large $Q'_{\star}$) is predicted for those around less significant convective zone stars. This reinforces the idea discussed above that large stellar convective envelopes seem to be the key to understand the quick tidal evolution of close-in planets. However, we caution that this trend is based on a small sample of stars with planets, for some of which only upper and lower limits of $Q'_{\star}$ are measured, therefore more data with actual values of the modified tidal quality factor are needed to confirm or discard these results.   

In this regard, space missions \textit{Kepler} \citep{borucki} and K2 \citep{howell14} represent an invaluable source of information. They have not only observed the largest collection of stars with transiting planets known so far, but also provided several transits per target. This implies a huge amount of potential empirical measurements of the modified tidal quality factor by precisely determining the mid-transit times of those targets observed during several consecutive years. These values combined with the already available estimations of effective temperatures for the host stars, can further fill out the  Figure \ref{teffvsq} and confirm the trend found in this study.

Finally, we did not find any significant correlation between the planetary mass, the ratio of the planet to the stellar mass, and the orbital period with $Q'_{\star}$. In Table \ref{table:9}, we list the main values of the planetary and stellar properties used to plot the Figure \ref{teffvsq}.

\begin{table*}
\renewcommand\thetable{9}
\caption{Planetary and stellar properties of the 15 systems used to test the predictions by Zahn}             
\label{table:9}      
\centering          
\begin{tabular}{ccccccccc}     
\hline\hline       
 System	&	$T_\mathrm{eff}$	 &	$\log g$ 	&	$P$[days]	&	$M_\mathrm{P}$ [$M_\mathrm{Jup}$]	& $Q'_{\star}$  &	M$_{cz}$/M$_{\star}$($\%$) 	&	Ref. $Q'_{\star}$	&	Ref $T_\mathrm{eff}$ and $\log g$\\
\hline                    
Kepler-78	&	5089	$\pm$	50	&	4.6	$\pm$	0.1	&	0.3550074	&	0.025	&	14695$^{b}$	&	9.624195	&	1	&	1	\\
WASP-19	&	5591	$\pm$	62	&	4.46	$\pm$	0.09	&	0.78884	&	1.114	&	1230000	&	3.609035	&	This work	&	2	\\
WASP-43	&	4400	$\pm$	200	&	4.5	$\pm$	0.2	&	0.813477	&	2.052	&	100000	&	21.734019	&	3	&	4	\\
WASP-103	&	6110	$\pm$	160	&	4.22$^{+0.12}_{-0.05}$	&	0.925542	&	1.49	&	1000000	&	0.245736	&	5	&	6	\\
WASP-18	&	6526	$\pm$	69	&	4.73	$\pm$	0.08	&	0.941451	&	10.43	&	1000000	&	0.132701	&	7	&	2	\\
KELT-16	&	6236	$\pm$	54	&	4.253$^{+0.031}_{-0.036}$	&	0.968995	&	2.75	&	110000	&	0.215928	&	5	&	8	\\
WASP-12	&	6313	$\pm$	52	&	4.37	$\pm$	0.12	&	1.0914203	&	1.47	&	182000	&	0.010152	&	5	&	2	\\
HAT-P-23	&	6000	$\pm$	125	&	4.5	$\pm$	0.2	&	1.212884	&	2.09	&	560000	&	1.088569	&	5	&	9	\\
KELT-1	&	6516	$\pm$	49	&	4.234$^{+0.012}_{-0.018}$	&	1.217514	&	27.23	&	840000	&	0.021936	&	5	&	10	\\
WASP-33	&	7430	$\pm$	100	&	4.3	$\pm$	0.2	&	1.219869	&	2.1	&	880000	&	0.000004	&	5	&	11	\\
WASP-4	&	5513	$\pm$	43	&	4.5	$\pm$	0.1	&	1.338231	&	1.237	&	29000	&	4.580076	&	12	&	2	\\
WASP-46	&	5761	$\pm$	16	&	4.47	$\pm$	0.06	&	1.43037	&	2.101	&	7000	&	1.654130	&	13	&	13	\\
Kepler-1658	&	6216	$\pm$	78	&	3.673	$\pm$	0.026	&	3.8494	&	5.88	&	121900	&	0.000894	&	14	&	14	\\
XO-1	&	5754	$\pm$	42	&	4.61	$\pm$	0.05	&	3.941512	&	0.9	&	400000	&	2.745198	&	15	& 2	\\
K2-39	&	4881	$\pm$	20	&	3.44	$\pm$	0.07	&	4.60543	&	0.158	&	0.178$^{b}$	&	46	&	16	&	16\\
\hline                  
\end{tabular}

References: (1) \citet{sanchis13}; (2) \citet{mortier}; (3) \citet{hoyer}; (4) \citet{hellierw43}; (5) \citet{macie}; (6) \citet{gillon}; (7) \citet{wilkins}; (8) \citet{oberst}; (9) \citet{bakos}; (10) \citet{siverd}; (11) \citet{collier}; (12) \citet{bouma}; (13) \citet{petrucci18}; (14) \citet{chontos}; (15) \citet{southworth}; (16) \citet{vaneylen}.

$^{b}$: This value of $Q'_{\star}$ was computed in this work by replacing in Eq. \ref{coeficiente_tidal} the measurements of $dP/dt$ and those of the planetary and stellar properties published in the paper cited in the seventh column, where $dP/dt$ was obtained by fitting a quadratic model to the mid-transit times as a function of epoch.   
\end{table*}

\section*{Acknowledgements}

This work has been partially supported by UNAM-PAPIIT IN-107518. R. P. thanks Carl Knight and Ana\"{e}l W\"{u}nsche for nicely providing information about the observations of WASP-19b published in the ETD. R. P. and E. J. acknowledge DGAPA for their postdoctoral fellowships and Drs. L. Hebb, C. Hellier, J. Tregloan-Reed, D. Dragomir, J. Bean, E. Sedaghati and N. Espinoza for kindly providing the new transits presented in their studies. R. P. and E. J. are also grateful to the operators of the 1.54-m telescope at EABA, Cecilia Qui\~{n}ones and Luis Tapia, for their support during the observing runs. All the authors thank the anonymous referee for useful comments and suggestions. We would also like to thank the Pierre Auger Collaboration for the use of its facilities. The operation of the robotic telescope FRAM is supported by the grant of the Ministry of Education of the Czech Republic LM2015038. The data calibration and analysis related to FRAM telescope is supported by the Ministry of Education of the Czech Republic MSMT-CR LTT18004 and MSMT/EU funds CZ.02.1.01/0.0/0.0/16$\_$013/0001402. This research has made use of the SIMBAD database, operated at CDS, Strasbourg, France and NASA's Astrophysics Data System Bibliographic Services.




\bibliographystyle{mnras}
\bibliography{wasp19.bib} 








\bsp	
\label{lastpage}
\end{document}